\documentclass[ams,aps,pre,twocolumn,groupedaddress,showpacs,floatfix]{revtex4}
\usepackage{amssymb}
\usepackage{graphicx}
\usepackage{amsmath}

\begin{document}

\title{The Analytical Theory of Bulk Melting II: \\
Variational Method Solution in the FCC Crystal}
\author{Yajun Zhou \footnote{%
Present address: Department of Chemistry and Chemical Biology,
Harvard
University, Cambridge, MA 02138, USA.} and Xiaofeng Jin \footnote{%
To whom correspondence should be addressed. Email: xfjin@fudan.ac.cn}}
\affiliation{Surface Physics Laboratory \& Department of Physics, Fudan University,
Shanghai 200433, China}
\date{\today}

\begin{abstract}
Continuing the arguments in Paper~I (arXiv: cond-mat/0405487), we model the
temperature dependence of interstitial defects in a surface-free
face-centered-cubic (fcc) elemental crystal and obtain the free energy and
correlation behavior based on variational methods. We show that the
avalanche of interstitial defects is the instability mechanism at the
melting point that bridges Lindemann and Born criteria.
\end{abstract}

\maketitle

\section{Introduction}

Bulk melting is a solid-liquid phase transition (SLPT) taking place in a
surface-free crystal \cite{Cahn}. The study of this phenomenon helps to
clarify the essential driving force of the inhomogeneous phase transition
out of a homogeneous system in that the surface-free condition rules out the
influence of inhomogeneity at the surface boundary. Many endeavors have been
made to test the previous melting theories by probing into the behavior of
bulk melting \cite{rev}, especially in search of the relation between the
widely-cited Lindemann \cite{Lindemann} and Born \cite{Born} criteria.
Lindemann criterion proposes that melting is triggered by the avalanche of
the root-mean-square~(rms) atom displacement after it exceeds a threshold
fraction~($\delta _{L}^{\ast }$) of the atom spacing~($a$), where $\delta
_{L}^{\ast }$ is called the critical Lindemann ratio, a semi-empirical
parameter once conceived as a lattice type characteristic; Born criterion
argues that the vanishing of the shear modulus is responsible for the
inability to resist lattice destruction at the melting point.

In the recent molecular dynamics simulation contributed by Jin \textit{et al.%
}~\cite{Jin}, it is demonstrated that the melting point of a
surface-free Ar crystal does satisfy both Lindemann and Born
criteria. In Ref.~\cite{Jin}, the Ar crystal melts when the shear
moduli see a sudden downfall (albeit not vanishing, which is
consistent with the experimental observation of residual shear
modulus at the melting point \cite{Brule}) and the atom
displacement surges towards infinity simultaneously. Jin
\textit{et al.}~have attributed this coincidence to the
\textquotedblleft Lindemann particle\textquotedblright\ (atom with
displacement larger than $\delta _{L}^{\ast }$ times the atom
spacing) clusters that emerge sporadically at low temperatures but
abound throughout the crystal when melting point is approached. It
is found that within such clusters (\textquotedblleft liquid
nuclei\textquotedblright ) consisting of energetic
\textquotedblleft Lindemann particles\textquotedblright , the
shear moduli difference is nearly
vanishing: $\Delta C_{S}=C_{44}-(C_{11}-C_{12})/2\approx 0$. Judging the r%
\^{o}le of \textquotedblleft Lindemann particle
clusters\textquotedblright\ that satisfy both the Born and
Lindemann instability criteria, it is then concluded in
Ref.~\cite{Jin} that melting is governed by the \textquotedblleft
strongly-correlated\textquotedblright\ Lindemann and Born
perspectives. However, several important questions remain
unanswered by numerical results:

\noindent \emph{Question 1:}~What kind of instability mechanism is directly
responsible for the non-zero shear moduli at the melting point? Is that
contradictory to Born's original argument that melting is triggered by zero
shear moduli?

\noindent \emph{Question 2:}~How does a heterogeneous nucleation process
manage to come out from a homogeneous system? Does the \emph{finite}-size of
the liquid nucleus at the melting point violate the principle
\textquotedblleft phase transition in the thermodynamic
limit\textquotedblright , which requires a system consisting of infinitely
many atoms?

\noindent \emph{Question 3:}~What are the most important factors, in terms
of the parameters of the\ interatomic forces, that affect the magnitude of
the parameter $\delta _{L}^{\ast }$? How general is this $\delta _{L}^{\ast
} $ \cite{Lrule}?

In this paper serving as an extension of the model in Paper~I to
the three-dimensional (3D) face-centered-cubic (fcc) elemental
crystal, we report the detailed analytical procedure that joins
the Lindemann and Born criteria together in the SLPT and the
mathematical arguments that address the three numbered questions
above. We show that the spatial correlation between interstitial
defects is the impetus that triggers and propagates instability in
a crystal and that eventually undermines long-range order in the
surface-free solid. Following the idea of defect-motivated
transition in previous literature \cite{surface melting, Stilinger
and Weber, first-order transition, Granato} and the previous
understandings of defects' cooperation and aggregation
\cite{Baskes, Halpern, Sandberg, Self-interstitial,
Si-interstitial,
Ge-interstitial, As-interstitial, Stilinger and Weber, Juelich}, we use the $%
J_{1}$-$J_{2}$ lattice model plus vibrational Hamiltonian \cite{deffreq} to
formulate the cooperative creation and the spatial correlation of
interstitials. Our results, which are based on fewer empirical parameters
than that in Ref.~\cite{Granato}, not only epitomize the origin of
instability and atom-scale pathway in the melting process with mathematical
rigor but also help to elucidate the origin of \textquotedblleft Lindemann
particle clusters\textquotedblright\ and to explain the experimental fact
\cite{Lrule} that $\delta _{L}^{\ast }$ is modulated by both lattice
geometry and the profile of interactions.

Paper~II is organized as follows: Sect.~II is a brief account of
the methodology of our model for interstitial defects with a
discussion of the symmetry of the Hamiltonian; Sect.~III uses
variational method to work out a mean field approximation (MFA)
solution to the three-dimensional (3D) model, where the
temperature dependence of the concentration and the correlation of
interstitial defects are investigated; Sect.~IV discusses the
physical implications of the model solutions where analytical
results are compared to data from experiments and simulation;
Sect.~V summarizes the results in Paper~II. Appendix~A provides an
alternative derivation of the formula which describes the
temperature dependence of concentration of defect; Appendix~B
gives the mathematical details involved in the evaluation of the
correlation of defects.

\section{Model Hamiltonian and its Qualitative Properties}

\subsection{Total Hamiltonian}

We begin our argument with the Hamiltonian describing a system of $N=4\ell
^{3}$ atoms, labeled as $\alpha =1,\dots ,N$:
\begin{eqnarray}
\mathcal{H} &\mathcal{=}&\sum_{\alpha =1}^{N}T_{\alpha }+\frac{1}{2}%
\sum_{\alpha =1}^{N}\sum_{\beta =1}^{N}V_{\alpha \beta }  \notag \\
&=&\sum_{\alpha =1}^{N}T_{\alpha }+\frac{1}{2}\sum_{\alpha
=1}^{N}\sum_{\beta =1}^{N}\left( V_{\alpha \beta }^{\text{vib}}+V_{\alpha
\beta }^{\text{conf}}\right)  \notag \\
&=&%
\begin{array}{c}
\\
\underbrace{\sum_{\alpha =1}^{N}\left( T_{\alpha }+\frac{1}{2}\sum_{\beta
=1}^{N}V_{\alpha \beta }^{\text{vib}}\right) } \\
\mathcal{H}^{\text{vib}}%
\end{array}%
+%
\begin{array}{c}
\\
\underbrace{\frac{1}{2}\sum_{\alpha =1}^{N}\sum_{\beta =1}^{N}V_{\alpha
\beta }^{\text{conf}}} \\
\mathcal{H}^{\text{conf}}%
\end{array}%
.  \label{H}
\end{eqnarray}%
This Hamiltonian incorporates the kinetic energy of each atom labeled $%
\alpha $~($T_{\alpha }$) and pairwise potential energy between atoms $\alpha
$ and $\beta $~($V_{\alpha \beta }$), where $V_{\alpha \alpha }=0$, for $%
\alpha =1,\dots ,N$. For each atom labeled $\alpha $, the potential energy
which it experiences is the summation of the contribution from all the other
atoms, which reads $\sum_{\beta =1}^{N}V_{\alpha \beta }$. From this sum, we
extract a \textquotedblleft vibrational potential\textquotedblright\ $%
\sum_{\beta =1}^{N}V_{\alpha \beta }^{\text{vib}}$, which is harmonically
oscillating with respect to interatomic distance when the atom $\alpha $ is
\emph{perturbed}. The \textquotedblleft configurational
potential\textquotedblright\ $\sum_{\beta =1}^{N}V_{\alpha \beta }^{\text{%
conf}}$ is formally defined as $\sum_{\beta =1}^{N}(V_{\alpha \beta
}-V_{\alpha \beta }^{\text{vib}})$. This Hamiltonian separation method has
been employed to study the lattice vibration's influence on the
concentration of crystal defects \cite{deffreq}.

The model in Paper~II basically treats an elemental crystal with fcc
structure~(Ar, for instance). In order to take the interstitial defects into
account, we use a lattice model with the standard NaCl-type structure. At
absolute zero, the Na-like lattice is totally occupied by Ar atoms, while
the Cl-like lattice is vacant and forms the totality of octahedral holes in
a perfect fcc crystal. Mathematically speaking, if one denotes a lattice
position $\mathbf{r}$ by a triad $\mathbf{(}hkl\mathbf{)}$, the Na-like (%
\textit{or }Cl-like) lattice is characterized by odd (\textit{or }even) $%
h+k+l$ indices. At any finite temperatures, by disregarding the vibrations
and lattice distortions (that is, by neglecting the atoms' kinetic energy
and its consequences) momentarily \cite{vol}, one may caricature the motion
of the atoms in the solid as the stochastic hops on the two interpenetrating
fcc lattices. The configurations of the system are thus exhausted by all the
possible ways to arrange $N$ atoms of a fcc elemental crystal at $2N$
possible sites, so the summation over atom-pairs could be converted to a
summation over site-pairs, as long as the site occupancy rate $n$ at
position $\mathbf{r=(}hkl\mathbf{)}$ is included in the following way:
\begin{eqnarray}
&&\frac{1}{2}\sum_{\alpha =1}^{N}\sum_{\beta =1}^{N}V_{\alpha \beta }^{\text{%
conf}}=\frac{1}{2}\sum_{\mathbf{r}}\sum_{{\mathbf{r}^{\prime }}}J_{\mathbf{r}%
{\mathbf{r}^{\prime }}}^{\text{conf}}n_{\mathbf{r}}n_{\mathbf{r}^{\prime }}
\notag \\
&\approx& \frac{1}{2}\sum_{h,k,l=1}^{2\ell }n_{hkl}\Bigg( %
J_{1}\sum_{\left\vert h-h^{\prime }\right\vert +\left\vert k-k^{\prime
}\right\vert +\left\vert l-l^{\prime }\right\vert =1}n_{h^{\prime }k^{\prime
}l^{\prime }}  \notag \\
&&+J_{2}\sum_{\left\vert h-h^{\prime \prime }\right\vert +\left\vert
k-k^{\prime \prime }\right\vert +\left\vert l-l^{\prime \prime }\right\vert
=2}n_{h^{\prime \prime }k^{\prime \prime }l^{\prime \prime }}\Bigg)
\label{Hconf}
\end{eqnarray}%
Here, $\sum_{\mathbf{r}}$ denotes summation over all the $2N$ sites, and $n_{%
\mathbf{r}}$ denotes the number of the atoms occupying the site at position $%
\mathbf{r}$, and $n_{\mathbf{r}}=0$ or $1$. The \textquotedblleft coupling
constant\textquotedblright\ $J_{\mathbf{r}\mathbf{r}^{\prime }}^{\text{conf}%
} $ in Eqn.~(\ref{Hconf}) equals to $V_{\alpha \beta }^{\text{conf}}$ when
the distance between two atoms $\alpha $ and $\beta $ is $\left\vert \mathbf{%
r}-\mathbf{r}^{\prime }\right\vert $. We have used the following cutoff in
Eqn.~(\ref{Hconf}): $J_{\mathbf{r}\mathbf{r}^{\prime }}^{\text{conf}}=J_{1}$
when $\left\vert \mathbf{r}-\mathbf{r}^{\prime }\right\vert =$ the
nearest-neighbor~(NN) distance, namely, the distance between a pair of
nearest \textquotedblleft Na\textquotedblright\ and \textquotedblleft
Cl\textquotedblright ; $J_{\mathbf{r}\mathbf{r}^{\prime }}^{\text{conf}%
}=J_{2}<0$ when $\left\vert \mathbf{r}-\mathbf{r}^{\prime }\right\vert =$
the next-nearest-neighbor~(NNN) distance, namely, the distance between a
pair of nearest \textquotedblleft Na\textquotedblright\ sites (or
\textquotedblleft Cl\textquotedblright\ sites); $J_{\mathbf{r}{\mathbf{r}%
^{\prime }}}^{\text{conf}}=0$ otherwise. In order that the NNN distance
becomes the bond length in the fcc crystal in our model (such as Ar, as
opposed to NaCl), we require that $J_{1}>J_{2}$ and $J_{2}<0$. This $J_{1}$-$%
J_{2}$ lattice model approximation is justified for interatomic forces that
are both pairwise and short-ranged, which is physically applicable to solids
where the interaction is governed by the Lennard-Jones (as in noble gases)
or Morse functions (as in some metals) \cite{Juelich}.

\begin{figure}[t]
\center{\includegraphics[width=8cm]{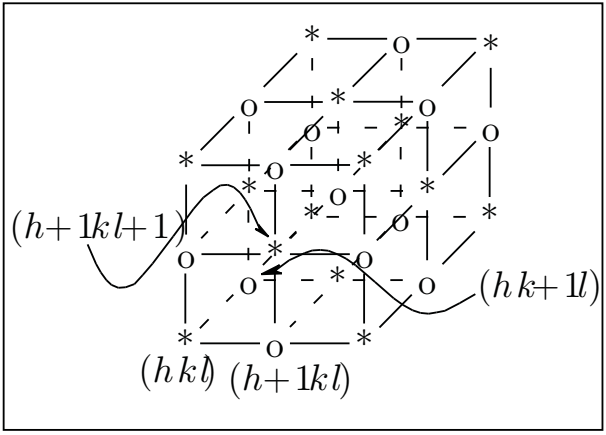}} \caption{The way we
label the lattice sites (``$*$'') and interstitial sites (``o'').}
\label{FIG1:a}
\end{figure}

Therefore, in the fcc crystal with $2N=8\ell ^{3}$ sites and $N=4\ell ^{3}$
atoms (FIG.~\ref{FIG1:a}), the configurational Hamiltonian now reads

\begin{eqnarray}
\mathcal{H}^{\text{conf}} &=&\frac{1}{2}\sum_{h,k,l=1}^{2\ell }n_{hkl}\Bigg( %
J_{1}\sum_{\left\vert h-h^{\prime }\right\vert +\left\vert k-k^{\prime
}\right\vert +\left\vert l-l^{\prime }\right\vert =1}n_{h^{\prime }k^{\prime
}l^{\prime }}  \notag \\
&&+J_{2}\sum_{\left\vert h-h^{\prime \prime }\right\vert +\left\vert
k-k^{\prime \prime }\right\vert +\left\vert l-l^{\prime \prime }\right\vert
=2}n_{h^{\prime \prime }k^{\prime \prime }l^{\prime \prime }}\Bigg)
\label{fccH}
\end{eqnarray}%
with the cyclic boundary condition:
\begin{equation}
n_{h+2\ell ,k,l}=n_{h,k+2\ell ,l}=n_{h,k,l+2\ell }=n_{h,k,l}
\end{equation}%
that eliminates the surface and the atom number conservation condition:
\begin{equation}
\sum_{h,k,l=1}^{2\ell }n_{hkl}=N
\end{equation}%
that reduces the degree of independence for occupancy rate by one.
Similar to the nomenclature in Paper~I, we call the sites bearing
the label $\left( hkl\right) $ where $h+k+l$ is odd (\textit{or
}even), namely, \textquotedblleft Na\textquotedblright\
(\textit{or }\textquotedblleft Cl\textquotedblright ) sites, as
lattice (\textit{or } interstitial) sites, or vice versa. The
lattice sites and interstitial sites interpenetrate, as they did
in the one-dimensional (1D) model in Paper~I.

In the 3D crystal, $\mathcal{H}^{\text{vib}}$ is taken into consideration in
the form of vibrational free energy
\begin{equation}
F^{\text{vib}}=3Nk_{B}T\log \frac{h\left\langle \nu \right\rangle }{k_{B}T},
\end{equation}%
where $k_{B}$ is the Boltzmann constant, $T$ is the absolute temperature, $h$
is the Planck constant, $\left\langle \nu \right\rangle $ is the geometrical
mean frequency of the crystal \cite{Landau, deffreq}. It should be noticed
that the formula above is valid in the Dulong-Petit limit -- the temperature
region where most melting processes take place and the heat capacity
contributed by lattice vibration is asymptotically $3Nk_{B}$.

\subsection{Symmetry of the Configurational Hamiltonian}

We now study the symmetry properties of the configurational Hamiltonian
under the transformation: $J_{1}\mapsto -J_{1}$. This will shed light on the
properties of the \textquotedblleft liquid phase\textquotedblright\ derived
from our model Hamiltonian as well as the interplay of $J_{1}$ and $J_{2}$.

First, by using the transformation $\sigma _{hkl}=2n_{hkl}-1$ \cite{Pathria,
Huang, Yang and Lee}, one could map the model Hamiltonian in Eqn.~(\ref{fccH}%
) to the $J_{1}$-$J_{2}$ model for magnetism \cite{Frustrated magnets},
where $\sigma _{hkl}=\pm 1$:

\begin{eqnarray*}
&&\mathcal{H}^{\text{conf}}\left( J_{1},J_{2},\{\sigma \}\right)- \frac{3N}{2%
}\left(J_{1}+2J_{2}\right) \\
&=&\frac{1}{8}\sum_{h,k,l=1}^{2\ell }\sigma _{hkl}\Bigg(J_{1}\sum_{\left%
\vert h-h^{\prime }\right\vert +\left\vert k-k^{\prime }\right\vert
+\left\vert l-l^{\prime }\right\vert =1}\sigma _{h^{\prime }k^{\prime
}l^{\prime }} \\
&&+J_{2}\sum_{\left\vert h-h^{\prime \prime }\right\vert +\left\vert
k-k^{\prime \prime }\right\vert +\left\vert l-l^{\prime \prime }\right\vert
=2}\sigma _{h^{\prime \prime }k^{\prime \prime }l^{\prime \prime }}\Bigg) ,
\end{eqnarray*}%
\begin{equation}
\sigma _{h+2\ell ,k,l}=\sigma _{h,k+2\ell ,l}=\sigma _{h,k,l+2\ell }=\sigma
_{h,k,l},\sum_{h,k,l=1}^{2\ell }\sigma _{h,k,l}=0.
\end{equation}%
The partition function corresponding to this configurational Hamiltonian
reads: $Q^{\text{conf}}\left( J_{1},J_{2},T\right) =\sum_{\{\sigma \}}\exp
\left( -\mathcal{H}^{\text{conf}}\left( J_{1},J_{2},\left\{ \sigma \right\}
\right) /k_{B}T\right) $.

Second, we notice that the Hamiltonian is invariant under any
transformations that swap interstitial sites and lattice sites,
\textit{i.e.}~mappings such as $\left( hkl\right) \mapsto \left(
h+1kl\right) $, $\left( hkl\right) \mapsto \left( hk+1l\right) $,
$\left( hkl\right) \mapsto \left( hkl+1\right) $. (This is called
``sublattice symmetry'' \cite{Nigel Goldenfeld}.) It is
conventional to assume \cite{Nigel Goldenfeld} that there exists a
temperature $T^{\prime }\geq 0$K, above which the partition
functions $Q^{\text{conf}}\left( J_{1},J_{2},T\right) $ and
$Q^{\text{conf}}\left( -J_{1},J_{2},T\right) $ both preserve the
\textquotedblleft sublattice symmetry\textquotedblright\ of the
Hamiltonian. In other words, when $T>T^{\prime }$, both the
$\left( J_{1},J_{2}\right) $ and $\left( -J_{1},J_{2}\right) $
systems are characterized by equal amount of atoms occupying the
lattice sites and interstitial sites. Since the number of atoms
occupying either type of sites
is $N/2$, we may infer that both partition functions (\textit{i.e.}~$Q^{%
\text{conf}}\left( J_{1},J_{2},T\right) $ and $Q^{\text{conf}}\left(
-J_{1},J_{2},T\right) $) are dominated by terms with the configuration $%
\left\{ \sigma \right\} $ satisfying:
\begin{equation}
\sum_{\substack{ h,k,l=1  \\ \left( -1\right) ^{h+k+l}=1}}^{2\ell }\sigma
_{hkl}=\sum_{\substack{ h^{\prime },k^{\prime },l^{\prime }=1  \\ \left(
-1\right) ^{h^{\prime }+k^{\prime }+l^{\prime }}=-1}}^{2\ell }\sigma
_{h^{\prime }k^{\prime }l^{\prime }}=0.  \label{halfN}
\end{equation}%
So the transformation
\begin{equation}
\sigma _{hkl}\mapsto \sigma _{hkl}^{\ast }=\left( -1\right) ^{h+k+l}\sigma
_{hkl}
\end{equation}%
sends one dominant configuration
\begin{equation}
\left\{ \sigma \right\} =\left\{ \sigma _{hkl},h,k,l\in \{1,2,\dots ,2\ell
\}\left\vert \sum_{h,k,l=1}^{2\ell }\sigma _{hkl}=0\right. \right\}
\end{equation}%
to another dominant configuration
\begin{equation}
\left\{ \sigma ^{\ast }\right\} =\left\{ \sigma _{hkl}^{\ast },h,k,l\in
\{1,2,\dots ,2\ell \}\left\vert \sum_{h,k,l=1}^{2\ell }\sigma _{hkl}^{\ast
}=0\right. \right\} \
\end{equation}%
in the partition function, and transforms the configurational Hamiltonian in
the following way:%
\begin{eqnarray}
\mathcal{H}^{\text{conf}}\left( J_{1},J_{2},\sigma \right) &\mapsto &%
\mathcal{H}^{\text{conf}}\left( J_{1},J_{2},\sigma ^{\ast }\right)  \notag \\
&=&\mathcal{H}^{\text{conf}}\left( -J_{1},J_{2},\sigma \right) +3NJ_{1},
\end{eqnarray}%
where
\begin{eqnarray}
&&\mathcal{H}^{\text{conf}}\left( J_{1},J_{2},\{\sigma ^{\ast }\}\right)-%
\frac{3N}{2}\left( J_{1}+2J_{2}\right)  \notag \\
&=&\frac{1}{8}\sum_{h,k,l=1}^{2\ell }\sigma _{hkl}^{\ast }\Bigg(%
J_{1}\sum_{\left\vert h-h^{\prime }\right\vert +\left\vert k-k^{\prime
}\right\vert +\left\vert l-l^{\prime }\right\vert =1}\sigma _{h^{\prime
}k^{\prime }l^{\prime }}^{\ast }  \notag \\
&&+J_{2}\sum_{\left\vert h-h^{\prime \prime }\right\vert +\left\vert
k-k^{\prime \prime }\right\vert +\left\vert l-l^{\prime \prime }\right\vert
=2}\sigma _{h^{\prime \prime }k^{\prime \prime }l^{\prime \prime }}^{\ast }%
\Bigg) .  \notag \\
&&
\end{eqnarray}%
From the obvious identity:%
\begin{eqnarray}
&&\sum_{\{\sigma \}}e ^{-\frac{\mathcal{H}^{\text{conf}}\left(
J_{1},J_{2},\left\{ \sigma \right\} \right) }{k_{B}T}}=\sum_{\left\{ \sigma
^{\ast }\right\} }e^{ -\frac{\mathcal{H}^{\text{conf}}\left(
J_{1},J_{2},\left\{ \sigma ^{\ast }\right\} \right) }{k_{B}T}},
\end{eqnarray}%
we see that when $T>T^{\prime }$, the configurational free energy ($F^{\text{%
conf}}\equiv -k_{B}T\log Q^{\text{conf}}$) has the following correspondence
due to \textquotedblleft sublattice symmetry\textquotedblright :%
\begin{eqnarray}
&&F^{\text{conf}}\left( J_{1},J_{2},T\right) -\frac{3N}{2}\left(
J_{1}+2J_{2}\right)  \notag \\
&=&F^{\text{conf}}\left( -J_{1},J_{2},T\right) -\frac{3N}{2}\left(
-J_{1}+2J_{2}\right) .
\end{eqnarray}%
This equation is parallel to Eqn.~(12) in Paper~I, which is a formula
invariant under the transformation $J_{1}\mapsto -J_{1}$.

The analysis above reveals that when $T>T^{\prime }$, the free energy of the
system (up to a ground energy term that reads $(3N/2)(\pm J_{1}+2J_{2})$ )
is sensitive to $J_{2}$ but probably insensitive to $J_{1}$ -- it is because
$F^{\text{conf}}\left( \pm J_{1},J_{2},T\right) -(3N/2)\left( \pm
J_{1}+2J_{2}\right) $ remains unchanged even when $J_{1}$ alters its sign,
manifesting the irrelevance of $J_{1}$ in this temperature region. This
scenario is consistent with the previous description of the liquid phase in
the \textquotedblleft lattice gas model\textquotedblright\ \cite{Yang and
Lee}, that is, the properties of the system is predominantly defined by the
NNN attractive bonding energy, and it is unnecessary to take the repulsion
at NN\ distance into account.

Therefore, we have obtained the \textquotedblleft liquid
phase\textquotedblright\ solution to our model, which highlights the
importance of $J_{2}$ at high temperature. In the following section, we will
show that for low temperatures $T\ll T^{\prime }$, our model behaves
differently as compared to the conclusions above, where different $J_{1}$
parameter could change the properties of $F^{\text{conf}}\left(
J_{1},J_{2},T\right) -3NJ_{1}/2$ drastically, indicating the loss of
sublattice symmetry of the free energy. We will show that at low
temperatures, the energy relationship $J_{1}>J_{2}<0$ binds atoms together
in the solid by breaking the \textquotedblleft sublattice
symmetry\textquotedblright ;\ near the melting point, the same energy
relationship helps to propagate instability in the system and undermines the
long-range order by revoking the \textquotedblleft sublattice
symmetry\textquotedblright .

\section{The variational approach to the 3D fcc crystal}

\subsection{Phenomenological Parameter Expansion of the Free Energy
Functional and the Model of Melting}

In our probe into the 3D fcc elemental crystal, we aim to obtain a solution
based on modified MFA. We first define a random variable called the local
order parameter
\begin{equation}
L\left( \mathbf{r}\right) =\left( -1\right) ^{h+k+l}\sigma _{hkl}=\left(
-1\right) ^{h+k+l}\left( 2n_{hkl}-1\right)
\end{equation}%
where site $\mathbf{r}$ bears the integer label $\left( hkl\right) $.
According to the Landau-Ginzburg theory of MFA \cite{Ginzburg}, the free
energy functional contributed by configurational Hamiltonian of the 3D fcc
model $\mathcal{H}^{\text{conf}}$ could be expanded phenomenologically as%
\begin{widetext}
\begin{eqnarray}
&&F^{\text{conf}}\left[ L\left( \mathbf{r}\right) \right] =-TS^{\text{conf}}\left[ L\left( \mathbf{r}\right) \right] +\frac{1}{2}\left(\frac{\sqrt{2}}{a}\right)^{3}%
\int \mathrm{d}^{3}\mathbf{r\ }\left\{ \left[ \left( 1-L^{2}\left( \mathbf{r}%
\right) \right) \mathcal{H}_{1}+\left( 1-L^{2}\left( \mathbf{r}\right)
\right) ^{2}\mathcal{H}_{2}\right] +\gamma \left[ \nabla L\left( \mathbf{r}%
\right) \right] ^{2}\right\}   \notag \\
&&
\end{eqnarray}
where $(\sqrt{2}/a)^{3}\int \mathrm{d}^{3}\mathbf{r\equiv }%
\sum_{\mathbf{r}}$ denotes the summation over all the $2N$ sites.
(Recall that the NN distance in the $2N$ sites is $a/\sqrt{2}$, so
the volume of the smallest cube on the lattice is $\left(
a/\sqrt{2}\right) ^{3}$, which forms the ratio between the
site summation $\sum_{\mathbf{r}}$ and the spatial integration $\int \mathrm{%
d}^{3}\mathbf{r}$.) $\mathcal{H}_{1}$ $(>0)$ and $\mathcal{H}_{2}$ $(<0)$
are two energy parameters to be determined later as functions of $J_{1}$ and
$J_{2}$. $\gamma $ $(>0)$ is the phenomenological \textquotedblleft domain
wall energy\textquotedblright\ coefficient. The configurational entropy $S^{%
\text{conf}}\left[ L\left( \mathbf{r}\right) \right] $, which appears in the
equation above, reads as the entropy of mixing \cite{Landau, Huang}:%
\begin{equation}
S^{\text{conf}}\left[ L\left( \mathbf{r}\right) \right] =-k_{B}\left( \frac{%
\sqrt{2}}{a}\right) ^{3}\int \mathrm{d}^{3}\mathbf{r}\mathbf{\ }\left[ \frac{%
1-L\left( \mathbf{r}\right) }{2}\log \frac{1-L\left( \mathbf{r}\right) }{2}+%
\frac{1+L\left( \mathbf{r}\right) }{2}\log \frac{1+L\left( \mathbf{r}\right)
}{2}\right] \text{.}
\end{equation}%
$F^{\text{vib}}\left[ L\left( \mathbf{r}\right) \right] $ is contributed by
vibrations with geometric mean frequencies $\left\langle \nu
_{l}\right\rangle $ (lattice mode) and $\left\langle \nu _{i}\right\rangle $
(interstitial mode) \cite{deffreq}, which reads
\begin{equation}
F^{\text{vib}}\left[ L\left( \mathbf{r}\right) \right] =\frac{3k_{B}T}{2}%
\left( \frac{\sqrt{2}}{a}\right) ^{3}\int \mathrm{d}^{3}\mathbf{r\ }\frac{%
\left( 1-L^{2}\left( \mathbf{r}\right) \right) }{4}\log \frac{\left\langle
\nu _{i}\right\rangle }{\left\langle \nu _{l}\right\rangle }+\mathrm{const}
\end{equation}%
when expanded in power series of $\left( 1-L^{2}\left(
\mathbf{r}\right) \right) $.

Now the total phenomenological free energy functional $F\left[ L\left(
\mathbf{r}\right) \right] $ reads:%
\begin{equation}
F\left[ L\left( \mathbf{r}\right) \right] =F^{\text{conf}}\left[ L\left(
\mathbf{r}\right) \right] +F^{\text{vib}}\left[ L\left( \mathbf{r}\right) %
\right] =\left( \frac{\sqrt{2}}{a}\right) ^{3}\int \mathrm{d}^{3}\mathbf{r\ }%
f\left( L\left( \mathbf{r}\right) ,\nabla L\left( \mathbf{r}\right) \right)
\end{equation}%
where

\begin{eqnarray}
f\left( L\left( \mathbf{r}\right) ,\nabla L\left( \mathbf{r}\right) \right)
&=&\left\{ \frac{1}{2}\left[ \left( 1-L^{2}\left( \mathbf{r}\right) \right)
\mathcal{H}_{1} +\left( 1-L^{2}\left( \mathbf{r}\right) \right) ^{2}%
\mathcal{H}_{2}-3k_{B}T(1-L^{2}\left( \mathbf{r}\right) )\Upsilon
\right]\right.
\notag \\
&&\left.+k_{B}T \left[ \frac{1-L(\mathbf{r})}{2}\log
\frac{1-L(\mathbf{r}) }{2} +\frac{1+L(\mathbf{r}) }{2}\log
\frac{1+L(\mathbf{r}) }{2}\right] +\frac{\gamma }{2}\left[ \nabla
L( \mathbf{r}) \right] ^{2} \right\} .
\end{eqnarray}%
\end{widetext}and $\Upsilon =(1/4)\log \left( \left\langle \nu _{l}\right\rangle
/\left\langle \nu _{i}\right\rangle \right) >0$ is a factor
modeling the \textquotedblleft lattice
softening\textquotedblright\ effect, that is, interstitials have a
lower vibration frequency than normal atoms at the lattice sites.
By variational methods, we may apply the Euler-Lagrange equation
\begin{equation}
\frac{\partial f\left( L\left( \mathbf{r}\right) ,\nabla L\left( \mathbf{r}%
\right) \right) }{\partial L\left( \mathbf{r}\right) }=\nabla \frac{\partial
f\left( L\left( \mathbf{r}\right) ,\nabla L\left( \mathbf{r}\right) \right)
}{\partial \nabla L\left( \mathbf{r}\right) },
\end{equation}%
to optimize the free energy functional with the boundary-free condition and
obtain the equation of motion for $L\left( \mathbf{r}\right) $ in the
\textquotedblleft Poisson equation\textquotedblright\ form:

\begin{equation}
\rho \left( L\left( \mathbf{r}\right) \right) =-\gamma \nabla ^{2}L\left(
\mathbf{r}\right) ,  \label{eqmotion}
\end{equation}%
where
\begin{eqnarray}
\rho \left( L\left( \mathbf{r}\right) \right) &=&\mathcal{H}_{1}L\left(
\mathbf{r}\right) -2\mathcal{H}_{2}L\left( \mathbf{r}\right) \left(
L^{2}\left( \mathbf{r}\right) -1\right)  \notag \\
&&-3k_{B}TL\left( \mathbf{r}\right) \Upsilon -k_{B}T\tanh ^{-1}L\left(
\mathbf{r}\right).
\end{eqnarray}

\begin{figure}[t]
\centerline{
\includegraphics[width=3in]{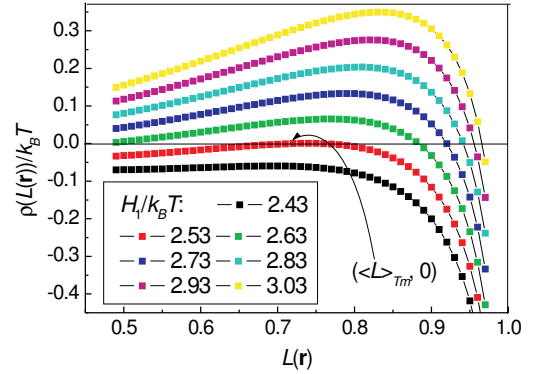}
}
\caption{ (color online) Local properties of $\protect\rho \left( L\left(
\mathbf{r}\right) \right) /k_{B}T$ near its node: $\langle L\mathfrak{%
\rangle }_{T}$. In this specific case, $\mathcal{H}_{1}=6\left\vert \mathcal{%
H}_{2}\right\vert $, $\Upsilon =\log \left( 4/3\right) .$ The tangent node $%
(\left\langle L\right\rangle _{T_{m}},0)$ is related to the melting point.}
\label{FIG2:b}
\end{figure}

The \textquotedblleft phenomenological electric charge
density\textquotedblright\ $\rho \left( L\left( \mathbf{r}\right) \right) $
immediately gives rise to a simple mathematical model of the catastrophe of
the long-range order: For sufficiently low temperature $T$, the curve $\rho
\left( L\left( \mathbf{r}\right) \right) $ intersects the positive $L\left(
\mathbf{r}\right) $-axis at least once (FIG.~\ref{FIG2:b}), so there is a
homogeneous distribution $L\left( \mathbf{r}\right) \equiv \langle L\rangle
_{T}\neq 0$ that makes $\rho \left( L\left( \mathbf{r}\right) \right) $
vanish. When $T=T_{m}$, where $T_{m}$ suffices
\begin{equation*}
\rho \left( \left\langle L\right\rangle _{T_{m}}\right) =0\text{ and }\left.
\frac{\partial \rho \left( L\left( \mathbf{r}\right) \right) }{\partial
L\left( \mathbf{r}\right) }\right\vert _{L\left( \mathbf{r}\right)
=\left\langle L\right\rangle _{T_{m}}}=0
\end{equation*}%
simultaneously, the intersection becomes a tangent point and we have the
following equations:
\begin{equation}
\left\{
\begin{array}{l}
3\Upsilon +\frac{1}{1-\left\langle L\right\rangle _{T_{m}}^{2}}=\frac{%
\mathcal{H}_{1}-2\mathcal{H}_{2}\left( 3\left\langle L\right\rangle
_{T_{m}}^{2}-1\right) }{k_{B}T_{m}} \\
\frac{1}{\left\langle L\right\rangle _{T_{m}}^{3}}\left( \frac{\left\langle
L\right\rangle _{T_{m}}}{1-\left\langle L\right\rangle _{T_{m}}^{2}}-\tanh
^{-1}\left\langle L\right\rangle _{T_{m}}\right) =-\frac{4\mathcal{H}_{2}}{%
k_{B}T_{m}}%
\end{array}%
\right. ,  \label{Tm}
\end{equation}
In FIG.~\ref{FIG2:b}, $k_{B}T_{m}=\mathcal{H}_{1}/2.53$. For temperatures
higher than $T_{m}$, $\rho \left( L\left( \mathbf{r}\right) \right) $ no
longer intersects the positive $L\left( \mathbf{r}\right) $-axis, suggesting
the loss of long-range order. From Eqn.~(\ref{Tm}) we know that as $\mathcal{%
H}_{2}$ tends to zero, so does $\left\langle L\right\rangle _{T_{m}}$.
Therefore, in the absence of \textquotedblleft virtual attraction between
defects\textquotedblright\ ($\propto \mathcal{H}_{2}$, to be elaborated
immediately in next two subsections), first-order SLPT is not possible
according to this model.

\subsection{The Cooperation Effect and the $\mathcal{H}_{2}$ Term}

In the MFA scenario,
\begin{eqnarray}
&&F\left[ L\left( \mathbf{r}\right) \right]\notag \\&=&F^{\text{vib}}-TS^{\text{conf}}+\frac{1}{2}\left( \frac{\sqrt{2}}{a}%
\right) ^{3}+\int \mathrm{d}^{3}\mathbf{r}\mathbf{\;\{(}6J_{1}\left\langle n_{%
\mathbf{r}}n_{\mathbf{r}^{\prime }}\right\rangle )\notag
\\ &&+12J_{2}\left\langle n_{\mathbf{r}}n_{\mathbf{r}^{\prime \prime
}}\right\rangle +\gamma \left[ \nabla L\left( \mathbf{r}\right)
\right] ^{2}\},  \label{FEfunctional}
\end{eqnarray}%
where $n_{\mathbf{r}}=\left( \sigma _{\mathbf{r}}+1\right) /2$,
$\mathbf{r}$
and $\mathbf{r}^{\prime }$ are NN sites, $\mathbf{r}$ and $\mathbf{r}%
^{\prime \prime }$ are NNN sites. One simple-minded approximation yields the
result: $\left\langle \sigma _{\mathbf{r}}\sigma _{\mathbf{r}^{\prime
}}\right\rangle =-\left\langle \sigma _{\mathbf{r}}\sigma _{\mathbf{r}%
^{\prime \prime }}\right\rangle =-L^{2}\left( \mathbf{r}\right) $.
This approximation is equivalent to the following statement: when
the correlation between sites is negligible and $L\left(
\mathbf{r}\right) $ distribution is uniform, the lattice
(\textit{or }interstitial) site occupancy is $\left(
1+\left\vert L\left( \mathbf{r}\right) \right\vert \right) /2$ (\textit{or }$%
\left( 1-\left\vert L\left( \mathbf{r}\right) \right\vert \right) /2$), so
that multiplication rule in probability theory infers that \cite{Huang},
\begin{eqnarray}
&&\left\langle n_{\mathbf{r}}n_{\mathbf{r}^{\prime }}\right\rangle \notag\\&=&\frac{%
1+\left\vert L\left( \mathbf{r}\right) \right\vert }{2}\frac{1-\left\vert
L\left( \mathbf{r}\right) \right\vert }{2}=\frac{1-L^{2}\left( \mathbf{r}%
\right) }{4},  \label{NNE} \\
&&\left\langle n_{\mathbf{r}}n_{\mathbf{r}^{\prime \prime
}}\right\rangle \notag\\&=& \frac{1}{2}\left( \left\langle
n_{\mathbf{r}}n_{\mathbf{r}^{\prime \prime
}}\right\rangle _{\text{lattice sites}}+\left\langle n_{\mathbf{r}}n_{%
\mathbf{r}^{\prime \prime }}\right\rangle _{\text{interstitial
sites}}\right) \notag\\&=&\frac{1}{2}\left[ \left(
\frac{1+\left\vert L\left( \mathbf{r}\right)
\right\vert }{2}\right) ^{2}+\left( \frac{1-\left\vert L\left( \mathbf{r}%
\right) \right\vert }{2}\right) ^{2}\right]\notag\\
&=&\frac{1}{2}-\frac{1-L^{2}\left( \mathbf{r}\right) }{4}.
\label{NNNE}
\end{eqnarray}
Such an argument is reasonable when $\left\vert L\left(
\mathbf{r}\right) \right\vert $ is sufficiently close to 1 and the
interstitial concentration
is low enough to obscure the cooperation between defects. When $\Upsilon =0$%
, the approximation above gives the defect concentration in the form of $%
\left( 1-\left\vert L\left( \mathbf{r}\right) \right\vert \right)
/2\sim \exp \left( -u/2k_{B}T\right) $ where $u=6J_{1}-12J_{2}$ is
the excitation energy of one interstitial defect. This asymptotic
behavior of the partition function agrees with the
well-established theory of Frenkel defects where
excitations of defects are assumed to be spatially independent \cite{deffreq}%
.

However, the approximation $\left\langle \sigma _{\mathbf{r}}\sigma _{%
\mathbf{r}^{\prime }}\right\rangle =-\left\langle \sigma _{\mathbf{r}}\sigma
_{\mathbf{r}^{\prime \prime }}\right\rangle $ is far from accurate when the
interstitial concentration is high and correlation between atoms should be
taken with care. By the qualitative analysis of the exact partition
function, we see from Eqn.~(\ref{halfN}) that $\left\langle \sigma _{\mathbf{%
r}}\sigma _{\mathbf{r}^{\prime }}\right\rangle $ vanishes in MFA scenario
when $T>T^{\prime }$, so that $F^{\text{conf}}\left( J_{1},J_{2},T\right)
-3NJ_{1}/2$ is independent from $J_{1}$ when $T>T^{\prime }$ while employing
MFA. However, it does not follow that $\left\langle \sigma _{\mathbf{r}%
}\sigma _{\mathbf{r}^{\prime \prime }}\right\rangle =-\left\langle \sigma _{%
\mathbf{r}}\sigma _{\mathbf{r}^{\prime }}\right\rangle =0$. We thus have to
make corrections to the estimate of $\left\langle \sigma _{\mathbf{r}}\sigma _{\mathbf{r}%
^{\prime \prime }}\right\rangle $, especially when $\left\vert
L\left( \mathbf{r}\right) \right\vert $ is far away from $1$. To
see the indispensability of this correction, we re-examine the
partition function
from the graph-theoretic perspective. In the thermodynamic limit, when $%
T>T^{\prime }$, by replacing the summation over all states with the
summation over thermodynamically most probable states (which is a
\textquotedblleft steepest-descent argument\textquotedblright\ similar to
Eqn.~(6)\ in Paper~I), we may write the exact partition function as:%
\begin{widetext}
\begin{eqnarray}
&&Q^{\text{conf}}\left( J_{1},J_{2},T\right) =\sum_{\{\sigma \}}\exp \left[ -%
\frac{\mathcal{H}^{\text{conf}}\left( J_{1},J_{2},\sigma \right) }{k_{B}T}%
\right] \rightarrow \sum_{\{\sigma _{hkl}=\pm 1\}}\exp \left[ -\frac{%
\mathcal{H}^{\text{conf}}\left( J_{1},J_{2},\sigma \right)
}{k_{B}T}\right]
\notag \\
&=&\frac{2^{2N}\exp \left[ -\frac{3N}{2k_{B}T}\left( J_{1}+2J_{2}\right)
\right] }{\cosh^{2N}\frac{J_{1}}{4k_{B}T}\cosh ^{2N}\frac{J_{2}}{4k_{B}T} }\sum_{r,s}g_{r,s}\tanh ^{r} \frac{J_{1}}{%
4k_{B}T} \tanh ^{s} \frac{J_{2}}{4k_{B}T} \notag \\
&\approx& \frac{%
2^{2N}\exp \left[ -\frac{3N}{2k_{B}T}\left( J_{1}+2J_{2}\right) \right] }{%
\cosh ^{2N} \frac{J_{2}}{4k_{B}T} }\sum_{s}g_{0,s}\tanh
^{s}\frac{J_{2}}{4k_{B}T}
\end{eqnarray}%
where $g_{r,s}$ is the number of loops that contain $r$ antibonds
(lines that join two NN sites) and $s$ bonds (lines that join two
NNN sites) and $r$ must be an even number because every loop is
closed. (The evenness of the number $r$ confirms again the
$J_{1}\mapsto -J_{1}$
symmetry. cf.~Appendix~A in Paper~I) The final \textquotedblleft $\approx $%
\textquotedblright\ in the equation above results from the fact that $\tanh
(J_{1}/4k_{B}T)\approx 0$, and that $\cosh (J_{1}/4k_{B}T)\approx 1$ since
the temperature $T$ is high enough. It follows that
\begin{eqnarray}
&&F^{\text{conf}}\left( J_{1},J_{2},T\right) -\frac{3N}{2}\left(
J_{1}+2J_{2}\right)  \notag \\
&=&-k_{B}T\log Q^{\text{conf}}\left( J_{1},J_{2},T\right) -\frac{3N}{2}%
\left( J_{1}+2J_{2}\right)  \notag \\
&=&2Nk_{B}T\log \cosh \frac{J_{2}}{4k_{B}T} -2Nk_{B}T\log 2
-k_{B}T\log \left( \sum_{s}g_{0,s}\tanh ^{s} \frac{J_{2}}{4k_{B}T}%
 \right)
\end{eqnarray}%
\end{widetext}still varies with respect to $T$ when $T>T^{\prime }$, and
this variation is dependent on the NNN attractive interaction $J_{2}$,
reinforcing that $\left\langle \sigma _{\mathbf{r}}\sigma _{\mathbf{r}%
^{\prime \prime }}\right\rangle $ is still non-zero when $T>T^{\prime }$.

From the behavior of the exact partition function, we see that the mean NNN
correlation is more persistent than its NN\ counterpart. In other words, the
system tends to \textquotedblleft preserve\textquotedblright\ more bonds
than assumed in the approximation $\left\langle \sigma _{\mathbf{r}}\sigma _{%
\mathbf{r}^{\prime }}\right\rangle =-\left\langle \sigma _{\mathbf{r}}\sigma
_{\mathbf{r}^{\prime \prime }}\right\rangle $. The inaccurate approximation $%
\left\langle \sigma _{\mathbf{r}}\sigma _{\mathbf{r}^{\prime }}\right\rangle
=-\left\langle \sigma _{\mathbf{r}}\sigma _{\mathbf{r}^{\prime \prime
}}\right\rangle $ infers that when interstitials are excited, the creation
of $6$ antibonds is accompanied by annihilation of exactly $12$ bonds, which
is not necessarily the case.

In order to offset the overestimate in the bond annihilation, we
have to write down
\begin{eqnarray}
&&6J_{1}\left\langle n_{\mathbf{r}}n_{\mathbf{r}^{\prime }}\right\rangle
+12J_{2}\left\langle n_{\mathbf{r}}n_{\mathbf{r}^{\prime \prime
}}\right\rangle  \notag \\
&=&6J_{2}+\left( 1-L^{2}\left( \mathbf{r}\right) \right) \mathcal{H}%
_{1}+\left( 1-L^{2}\left( \mathbf{r}\right) \right) ^{2}\mathcal{H}_{2}
\label{H1H2}
\end{eqnarray}%
where $\mathcal{H}_{1}=(3/2)\left( -2J_{2}+J_{1}\right) $, and the \textit{%
negative} coefficient $\mathcal{H}_{2}$, which is proportional to $J_{2}$,
is to be determined in the next subsection.

\subsection{The Phenomenological Parameters $\mathcal{H}_{2}$ and $\protect%
\gamma $ as Functions of $J_{1}$ and $J_{2}$}

\begin{figure}[tb]
\center{\includegraphics[width=8cm]{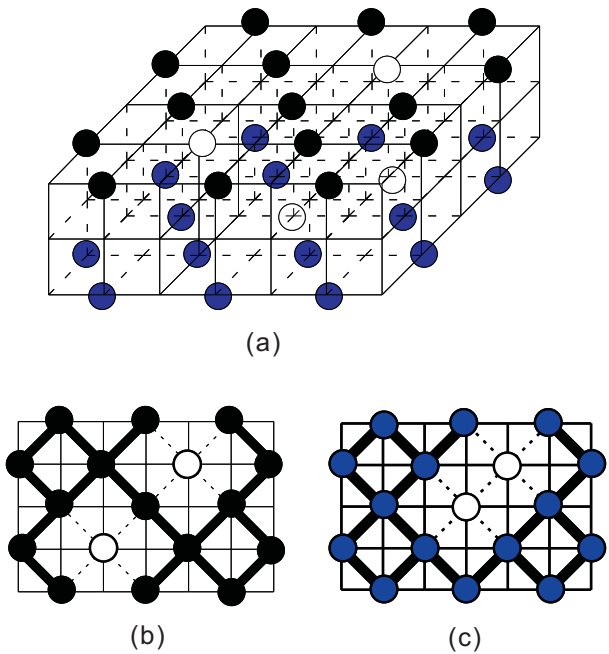}} \caption{(color
online) (a) This shows two layers in a crystal. In the
\textquotedblleft upper layer\textquotedblright , two holes are
separate whereas in the \textquotedblleft lower
layer\textquotedblright , two holes are close. (b) This shows the
cross-section view of the \textquotedblleft upper
layer\textquotedblright , with loss of eight bonds (in dashed
lines) as compared to a perfect layer. (c) This shows
cross-section view of the \textquotedblleft lower
layer\textquotedblright , with loss of seven bonds (in dashed
lines) as compared to a perfect layer. We could see that two close
holes save a bonding energy in the amount of $J_{2}$ (bonds are
denoted by thick black lines in (b) and (c)). } \label{FIG3:c}
\end{figure}

When the concentration of interstitials is considerable (about $3\%\sim 5\%$%
), the creation of di-interstitials (excitation of two interstitial atoms at
NNN distance) helps to preserve more NNN atom pairs than predicted in Eqn.~(%
\ref{NNNE}). This cooperation effect inherent in the Hamiltonian
lowers the energy cost in exciting interstitials because a
di-interstitial has less formation energy than that of two
interstitials at farther separate distance in our model
\cite{dumbbell}. Accordingly, we have to correct Eqn.~(\ref{NNNE})
by adding a positive term proportional to $( 1-L^{2}\left(
\mathbf{r}\right) ) ^{2}$ (symmetry and analyticity of the free
energy preclude terms including $L\left( \mathbf{r}\right) $ or
$\left\vert L\left( \mathbf{r}\right) \right\vert $). One scheme
to outline this correction is by estimating the percentage number
of occurrence of NNN-contact holes in the lattice sites as
\begin{equation}
\frac{12}{2}\left( \frac{1-\left\vert L\left( \mathbf{r}\right) \right\vert
}{2}\right) ^{2}\approx \frac{3}{8}\left( 1-L^{2}\left( \mathbf{r}\right)
\right) ^{2}
\end{equation}%
where \textquotedblleft $12$\textquotedblright\ is the coordination number
of NNN pairs. FIG.~\ref{FIG3:c} shows how \textquotedblleft virtual
attraction\textquotedblright\ \cite{Stilinger and Weber} between defects is
taken into consideration based on probabilistic arguments. The system
preserves more bonds than in the simple-minded approximation: when two holes
on the lattice sites are close (in NNN contact) instead of being farther
separate, they save the bond energy in the amount of $J_{2}$. In total, the
saved bonds by stochastic contact of holes lower the system energy in
magnitude of $\left( 3/8\right) \left\vert J_{2}\right\vert \left(
1-L^{2}\left( \mathbf{r}\right) \right) ^{2}$.

Dynamically speaking, this correction could be also understood by
the NNN\ correlation on the lattice sites. Since the occupancy of
lattice sites is approximately 1, the relaxation time of bonds on
lattice sites is much longer than that of the bonds on the
interstitial sites and the antibonds. So the dissociation of bonds
is a ``slower reaction'' than the creation of antibonds. That is
why $\left\langle n_{\mathbf{r}}n_{\mathbf{r}^{\prime
\prime }}\right\rangle _{\text{lattice sites}}$ should be greater than $%
\left[ \left( 1+\left| L\left( \mathbf{r}\right) \right| \right) /2\right]
^{2}$.

Therefore, after considering defect attraction or NNN\ correlation on the
lattice sites, the ensemble average $\left\langle n_{\mathbf{r}}n_{\mathbf{r}%
^{\prime \prime }}\right\rangle $ should be modified into
\begin{equation}
\left\langle n_{\mathbf{r}}n_{\mathbf{r}^{\prime \prime }}\right\rangle =%
\frac{1}{2}-\frac{1-L^{2}\left( \mathbf{r}\right) }{4}+\frac{3}{8}\left(
1-L^{2}\left( \mathbf{r}\right) \right) ^{2}\text{,}
\end{equation}%
so as to account for the crystal's tendency to retain as many NNN atom pairs
(bonds) as possible. Since NN sites are scarcely occupied by two atoms
simultaneously, NN\ correlations are immaterial as compared to NNN
correlations. In the light of this, we may retain Eqn.~(\ref{NNE}) in the
final expression of free energy functional. Comparing this with Eqn.~(\ref%
{H1H2}), we have
\begin{equation}
\mathcal{H}_{2}=\frac{3}{8}J_{2}<0.
\end{equation}

When spatial inhomogeneity is significant, a domain-wall energy proportional
to $\left[ \nabla L\left( \mathbf{r}\right) \right] ^{2}$ should be taken
into account in order to offset the miscalculation ($\Delta $) based on the
\textquotedblleft mean-field\textquotedblright\ assumption $L\left( \mathbf{r%
}\right) =L\left( \mathbf{r}^{\prime }\right) =L\left( \mathbf{r}^{\prime
\prime }\right) $. In Eqn.~(\ref{FEfunctional}), a good estimate of $%
\gamma $ is $a^{2}\mathcal{H}_{1}/2$ because

\begin{eqnarray}
&&\Delta \left( 2J_{1}\left\langle \sigma _{\mathbf{r}}\sigma _{\mathbf{r}%
^{\prime }}\right\rangle \right)  \notag \\
&=&2\bpm\includegraphics[height=0.3in]{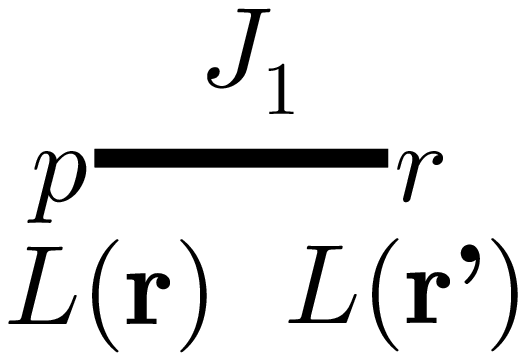}\epm-\bpm%
\includegraphics[height=0.3in]{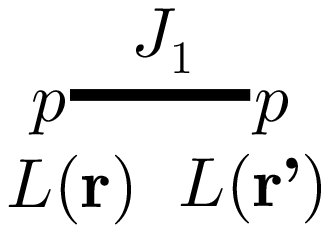}\epm-\bpm%
\includegraphics[height=0.3in]{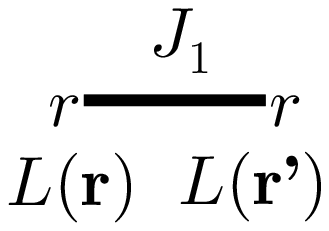}\epm  \notag \\
&=&-J_{1}\left( 2pr-p^{2}-r^{2}\right)  \notag \\
&=&J_{1}\left( p-r\right) ^{2}=J_{1}\frac{a^{2}}{2}\left[ \nabla L\left(
\mathbf{r}\right) \right] ^{2}
\end{eqnarray}%
\begin{eqnarray}
&&\Delta \left( 2J_{2}\left\langle \sigma _{\mathbf{r}}\sigma _{\mathbf{r}%
^{\prime \prime }}\right\rangle \right)  \notag \\
&=&2\bpm\includegraphics[height=0.4in]{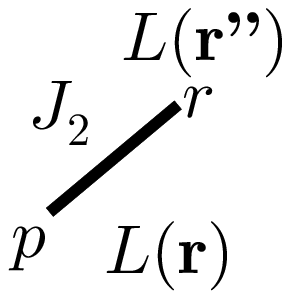}\epm-\bpm%
\includegraphics[height=0.4in]{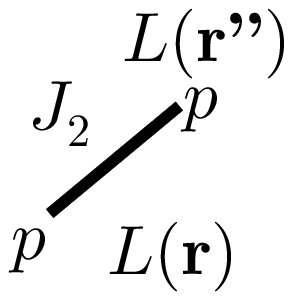}\epm-\bpm%
\includegraphics[height=0.4in]{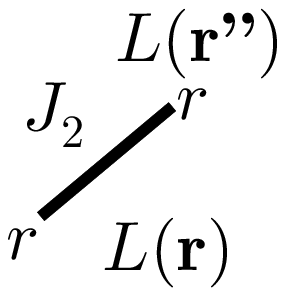}\epm  \notag \\
&=&-J_{2}a^{2}\left[ \nabla L\left( \mathbf{r}\right) \right] ^{2}.
\end{eqnarray}%
The total domain-wall energy contributed by the non-vanishing $\nabla
L\left( \mathbf{r}\right) $ at site $\mathbf{r}$ is thus
\begin{eqnarray}
&&\frac{1}{4}\times 6\times \frac{1}{2}\left( J_{1}\frac{a^{2}}{2}%
-J_{2}a^{2}\right) \left[ \nabla L\left( \mathbf{r}\right) \right] ^{2}
\notag \\
&=&\frac{a^{2}}{2}\mathcal{H}_{1}\left[ \nabla L\left( \mathbf{r}\right) %
\right] ^{2}.
\end{eqnarray}%
Here, the factor $1/4$ results from the transformation $\sigma _{\mathbf{r}%
}=2n_{\mathbf{r}}-1$, the factor $1/2$ is applied to offset repeated
counting, the number $6$ is a consequence of coordination number and the
Pythagorean theorem applied to the $\nabla L\left( \mathbf{r}\right) $
vector decomposition.

Physically speaking, at temperatures far lower than the Dulong-Petit limit,
the domain wall energy could be interpreted as the energy contributed by the
dislocations which destroy the continuity of $L\left( \mathbf{r}\right) $ on
the interface. The $(\gamma /2)\left[ \nabla L\left( \mathbf{r}\right) %
\right] ^{2}$ term associated with rare dislocations contribute little to
the total free energy at low temperatures. When temperature is sufficiently
high, large fluctuations of $L\left( \mathbf{r}\right) $ will be commonplace
and the gradient term will become indispensable if we want to evaluate the
total free energy correctly. (It should be emphasized that the vibration
term containing $\Upsilon $ arises from \textit{entropy} effect and does not
contribute to domain wall \textit{energy}. Even when the distribution of $%
L\left( \mathbf{r}\right) $ is inhomogeneous, the average vibrational
\textit{energy} for every degree of freedom is the same, say, $k_{B}T$,
regardless of vibrational frequency. Therefore, the expression of $\gamma $
is not dependent on $\Upsilon $.)

\subsection{\textquotedblleft Chemical Equilibrium\textquotedblright\ and
its Demise Due to Locally Isotropic Instability}

The physical implications behind the mathematical model outlined
in Sect.~III.A provide much more information about the melting
pathway.

First, for $T<T_{m}$, we find that $\rho \left( \left\langle L\right\rangle
_{T}\right) =0$ is equivalent to the \textquotedblleft chemical
equilibrium\textquotedblright\ condition:
\begin{figure}[b]
\center{\includegraphics[width=8cm]{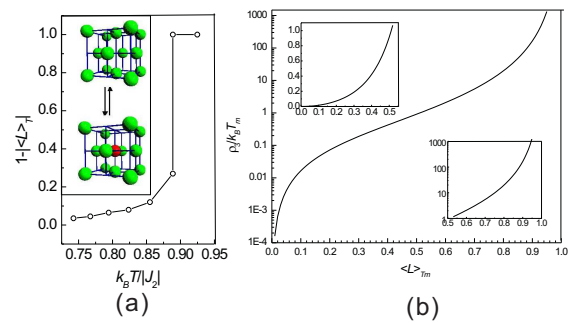}} \caption{(color
online) (a) This plots $\left\langle L\right\rangle _{T}-T$
dependence for the same $\mathcal{H}_{1}/|\mathcal{H}_{2}|$ and
$\Upsilon $ as in FIG.~\protect\ref{FIG2:b}, where catastrophe
occurs at the
\textquotedblleft melting point\textquotedblright\ $T_{m}=0.89|J_{2}|/k_{B}$%
. Inset shows a \textquotedblleft chemical equilibrium\textquotedblright\
between non-defective (up) and defective (down, interstitial defect
highlighted by red) cells at $T<T_{m}$. (b) Plotted here is the
dimensionless parameter $\protect\rho _{3}/k_{B}T_{m}$ as a function of $%
\left\langle L\right\rangle _{T_{m}}$ (inset shows details of the function
using different scales). $\protect\rho _{3}$ appears in the expression of $%
\protect\pi \protect\xi $, the critical size of the liquid nucleus.}
\label{FIG4:d}
\end{figure}
\begin{equation}
\frac{1-\left\vert \left\langle L\right\rangle _{T}\right\vert }{%
1+\left\vert \left\langle L\right\rangle _{T}\right\vert }=\frac{\left[
\text{defective cell}\right] }{\left[ \text{non-defective cell}\right] }=e^{-%
\frac{\Delta \overline{\varepsilon }}{k_{B}T}}.  \label{chem_eq}
\end{equation}%
Here, \textquotedblleft $\left[ \cdot \right] $\textquotedblright\ denotes
equilibrium concentration, and
\begin{equation}
\Delta \overline{\varepsilon }=2\left( \mathcal{H}_{1}-3k_{B}T\Upsilon
\right) \left\vert \left\langle L\right\rangle _{T}\right\vert -4\mathcal{H}%
_{2}\left\vert \left\langle L\right\rangle _{T}\right\vert (\left\vert
\left\langle L\right\rangle _{T}\right\vert ^{2}-1)
\end{equation}%
denotes the energy difference between two types of crystal cells: the
\textit{defective cell} incorporating (in the statistical parlance) more
than one (inclusive) interstitial and the \textit{non-defective cell}
including less than one interstitial (inset of FIG.~\ref{FIG4:d} (a)). (In
Appendix~A, we will derive the form of $\Delta \overline{\varepsilon }$ from
a perspective independent from the arguments in previous subsections.) This
dynamical equilibrium is achieved by stochastic creation and annihilation of
interstitials, or vividly speaking, the collisions in a dilute
\textquotedblleft gas\textquotedblright\ of interstitial monomers and
interstitial oligomers, in search of a minimal free energy corresponding to
an optimized long-range order parameter $\left\langle L\right\rangle _{T}$ \
(FIG.~\ref{FIG4:d} (a)).

Second, we notice that the aforementioned \textquotedblleft collision
process\textquotedblright\ causes a fluctuation of $L\left( \mathbf{r}%
\right) $, governed by Eqn.~(\ref{eqmotion}) that resembles the equation for
a globally neutral plasma when $T\ll T_{m}$. (This is because $\rho \left(
L\left( \mathbf{r}\right) \right) $ changes sign as $L\left( \mathbf{r}%
\right) $ varies in the vicinity of $\langle L\rangle _{T}$, which is
analogous to the coexistence of positive and negative charges in a plasma.)
The Green function of fluctuation response $G\left( \mathbf{r},\mathbf{r}%
^{\prime }\right) $ satisfies
\begin{equation}
\left( \frac{\sqrt{2}}{a}\right) ^{3}\left[ \gamma \nabla ^{2}+\frac{%
\partial \rho \left( L\left( \mathbf{r}\right) \right) }{\partial L\left(
\mathbf{r}\right) }\right] G\left( \mathbf{r},\mathbf{r}^{\prime }\right)
=-k_{B}T\delta \left( \mathbf{r}-\mathbf{r}^{\prime }\right) ,  \label{Green}
\end{equation}%
where $\delta \left( \mathbf{r}-\mathbf{r}^{\prime }\right) $ is the Dirac
delta function (See Appendix~B for the derivation of the Green function).
For sufficiently low temperature $T\ll T_{m}$, the partial derivative in the
equation above is always negative, so $G\left( \mathbf{r},\mathbf{r}^{\prime
}\right) $ is a propagator that decays exponentially to guarantee the
stability of the homogeneous distribution $L\left( \mathbf{r}\right) \equiv
\left\langle L\right\rangle _{T}$. However, as $T$ approaches $T_{m}$ from
below, it is possible that $\partial \rho \left( L\left( \mathbf{r}\right)
\right) /\partial L\left( \mathbf{r}\right) >0$ for certain values of $%
L\left( \mathbf{r}\right) $ not far from $\left\langle
L\right\rangle _{T}$, which results from a non-Gaussian
perturbation (Appendix~B) to the homogeneous distribution. In the
complex wave-number plane, apart from the poles at purely
imaginary wave-numbers (decay mode), the Fourier transform of the
Green function would then also encounter real wave-number
singularities (oscillation mode), indicating that homogeneity is
only preserved for a finite volume within which fluctuation
propagates as a sinusoidal wave. When $T=T_{m}$, which is defined
in Eqn.~(\ref{Tm}), the \textquotedblleft phenomenological
electric charge density\textquotedblright\ $\rho \left( L\left(
\mathbf{r}\right) \right) $ no longer changes its sign in the
vicinity of $\left\langle L\right\rangle _{T_{m}}$, and could be
expanded as $\rho \left( L\left( \mathbf{r}\right)
\right) =-\mu ^{2}\left( \mathbf{r}\right) \rho _{2}-\mu ^{3}\left( \mathbf{r%
}\right) \rho _{3}+\cdots $ in the vicinity of $\left\langle L\right\rangle
_{T_{m}}$, where $\mu \left( \mathbf{r}\right) =L\left( \mathbf{r}\right)
-\left\langle L\right\rangle _{T_{m}}$.

Physically speaking, this expansion of $\rho \left( L\left( \mathbf{r}%
\right) \right) $ infers that the uniform distribution $L\left( \mathbf{r}%
\right) \equiv \left\langle L\right\rangle _{T}$ \ is unstable when there is
a radial symmetric perturbation ($Y_{00}$ wave) of $L\left( \mathbf{r}%
\right) $ judging the Gauss' theorem:
\begin{eqnarray}
&&\gamma \oint_{S=\partial V}\nabla L\left( \mathbf{r}\right) \cdot \mathrm{d%
}\mathbf{S}  \notag \\
&=&-\int_{V}\rho \left( L\left( \mathbf{r}\right) \right) \mathrm{d}V\left\{
\begin{array}{lll}
>0, &  & L\left( \mathbf{r}\right) >0 \\
<0, &  & L\left( \mathbf{r}\right) <0%
\end{array}%
\right. .  \label{Gauss}
\end{eqnarray}%
When a perturbative $Y_{00}$ wave initiates, the surface integral boosts
when the spherical volume is increased, so the value of $L\left( \mathbf{r}%
\right) $ grows monotonously along the radial direction. Anisotropic $Y_{lm}$
modes do not apparently lead to steady growth of $L\left( \mathbf{r}\right) $
values and the related propagation of instability. It is because for $Y_{lm}$
modes where $l>0$, there is no such a causal relationship between the
surface integral's growth and the growth of the $L\left( \mathbf{r}\right) $
in the radial direction.

It can be verified mathematically that anisotropic $Y_{lm}$ modes are
quenched as $\sim r^{l}$ in the short range and do not account for the
instability. This is because $\mu \left( \mathbf{r}\right) $ satisfies:
\begin{equation}
\gamma \nabla ^{2}\mu \left( \mathbf{r}\right) =\mu ^{2}\left( \mathbf{r}%
\right) \rho _{2}+\mu ^{3}\left( \mathbf{r}\right) \rho _{3}.
\end{equation}%
For anisotropic excitations, where the asymptotic behavior is $\mu \left(
\mathbf{r}\right) \sim \mu \left( r\right) Y_{lm}\left( \theta ,\phi \right)
$, $r=\left\vert \mathbf{r}\right\vert \rightarrow 0$ (the original point of
$\mathbf{r}$ is placed at the center of excitation), $l>0$, we find
\begin{eqnarray}
\gamma \chi ^{\prime \prime }(r) &=&\frac{l(l+1)\gamma \chi (r)}{r^{2}}+%
\frac{\rho _{2}\chi ^{2}(r)Y_{lm}\left( \theta ,\phi \right) }{r}  \notag \\
&&+\frac{\rho _{3}\chi ^{3}(r)Y_{lm}^{2}\left( \theta ,\phi \right) }{r^{2}}
\notag \\
&\sim &\frac{l(l+1)\gamma \chi (r)}{r^{2}},
\end{eqnarray}%
where $\chi \left( \mathbf{r}\right) =\left\vert \mathbf{r}\right\vert \mu
\left( \mathbf{r}\right) $.
\begin{equation}
\mu \left( \mathbf{r}\right) \sim r^{l}\rightarrow 0,\text{ as }r\rightarrow
0
\end{equation}%
Meanwhile, $Y_{00}$ wave has the following asymptotic behavior:
\begin{equation}
\mu \left( \mathbf{r}\right) \sim -\frac{\rho _{2}}{\rho _{3}}\neq 0,\text{
as }r\rightarrow 0
\end{equation}%
Therefore, isotropic excitation dominates the instability mechanism at $%
T_{m} $, because all the $Y_{lm}$ ($l>0$) modes are overwhelmed by
the $Y_{00}$ mode in the short range.

\subsection{ Relations with Born and Lindemann Criteria: \textquotedblleft
Lindemann Particle Clusters\textquotedblright\ Revisited}

In the context of instability induced by locally isotropic excitations,
\textquotedblleft quasi-neutrality\textquotedblright\ in the
\textquotedblleft plasma\textquotedblright\ is attained by establishing $%
L\left( \mathbf{r}\right) >0$ and $L\left( \mathbf{r}\right) <0$ domains,
each in diameter of $\pi \xi \sim \sqrt{2}a\mathcal{H}_{1}^{2}\pi ^{3}(2\rho
_{3}k_{B}T_{m})^{-1}$ (Appendix~B elaborates on this estimate, and FIG.~%
\ref{FIG4:d}(b) plots the dimensionless parameter $\rho _{3}/k_{B}T_{m}$),
which is the average size of the locally isotropic excitations of atom
clusters. These highly cooperative and energetic atom clusters gives rise to
a catastrophe of global long-range order at $T_{m}$:
\begin{equation}
\left\vert \left\langle L\right\rangle _{T}\right\vert \left\{
\begin{array}{lll}
\rightarrow \left\vert \left\langle L\right\rangle _{T_{m}}\right\vert \neq
0, &  & T\rightarrow T_{m}-0 \\
=0, &  & T>T_{m}%
\end{array}%
\right. .
\end{equation}%
The vanishing long-range order $\left\vert \left\langle L\right\rangle
_{T}\right\vert $ at temperatures higher than $T_{m}$ casts the system into
sublattice symmetry.

Such locally isotropic excitations near $T_{m}$ form spherical
domains of instability (SDIs) and amount to one possible
interpretation for the origin of \textquotedblleft Lindemann
particles\textquotedblright\ \cite{Jin} and their aggregation. In
such SDIs, due to the nature of the fluctuation and the
relaxation, the displaced atoms are moving collaboratively,
energetically and isotropically, so that we could regard them as
the physical entity that bridges Born and Lindemann criteria at
the melting point according to the following argument: On one
hand, the totality of atoms in each SDI exhibits isotropy, which
guarantees exactly vanishing shear moduli difference: $\Delta
C_{S}=0$ -- that is, SDIs satisfy Born criterion; On the other
hand, the displacement of the atoms in all the SDIs exceeds
$a\delta _{L}^{\ast }$ -- that is, the average displacement of all
atoms in the Euclidean space, which packed spheres cannot
perfectly fill. In the light of this, the atoms in SDIs
are\textquotedblleft Lindemann particles\textquotedblright\ by
definition -- that is, SDIs also satisfy Lindemann criterion.

In short, the locally isotropic plasma instability at the superheating limit
(\textquotedblleft melting point\textquotedblright\ of a surface-free solid)
$T_{m}$ is initiated and propagated by SDIs, and $T_{m}$ is exactly the
temperature at which Born and Lindemann criteria coincide. These SDIs
declare the demise of the chemical equilibrium originally established by
collisions in a dilute \textquotedblleft gas\textquotedblright\ of
interstitial defects, and undermine the long-range order in a solid in the
mean time.

\section{Discussions}

From the arguments in the last section, we have provided accounts for the
instability mechanism (\textit{why}), molecular pathway (\textit{how}) and
phase transition temperature (\textit{when}) related to the melting of a
surface free fcc elemental crystal. The answers to \textit{why} and \textit{%
how} also help to clarify the physical basis for the equivalence of Born and
Lindemann criteria. This section will provide further implications of our
theoretical model and give answers to the numbered questions in the
Introduction.

The crucial instability mechanism (\textit{why}) of melting is
found to be the cooperative creation of interstitials that caused
the avalanche of displaced atoms. The cooperation effect transfers
the information of one displaced atom isotropically to its
neighborhood and finally results in the formation of SDIs. The
residual shear modulus at the melting point is due to the
inability of spherical domains to fully occupy the Euclidean
space. Born's argument is still valid within each SDI, but no
longer valid in the melting crystal as a whole. This answers
Question 1.

Therefore, for the whole crystal, lattice softening or vanishing
shear modulus is not as decisive in the melting process as the
cooperation effect is. In the melting point formula
Eqn.~(\ref{Tm}), it is clear that the first order transition
disappears when $\mathcal{H}_{2}$ is zero. The cooperation between
the interstitials proves to lower the energy cost to excite
defects thereby paving the way for destroying the long range order
in the lattice. For fixed bonding energy $J_{2}$ and lattice
softening $\Upsilon $, $T_{m}$
is lowered when $\left\vert \mathcal{H}_{2}\right\vert /\left\vert \mathcal{H%
}_{1}\right\vert $ ratio is enhanced, which is in accordance with the 1D
exact solution which relates high $\left\vert J_{2}\right\vert /\left\vert
J_{1}\right\vert $ ratio to great disorder.

To justify this, we will prove the following

\noindent \textit{Theorem:}
\begin{equation}
\left( \frac{\partial T_{m}}{\partial \mathcal{H}_{1}}\right) _{\mathcal{H}%
_{2},\Upsilon }>0
\end{equation}%
in the 3D fcc case.

\noindent \textit{Proof:} From Eqn.~(\ref{Tm}), pick $\left\langle
L\right\rangle _{T_{m}}>0$, we may arrive at the following conclusion:
\begin{widetext}
\begin{equation}
\frac{\partial }{\partial \left\langle L\right\rangle _{T_{m}}}\left( \frac{%
\mathcal{H}_{1}-3k_{B}T\Upsilon }{\left\vert \mathcal{H}_{2}\right\vert }%
\right) =-\frac{4\left\langle L\right\rangle _{T_{m}}\left( 1-\left\langle
L\right\rangle _{T_{m}}^{2}\right) ^{2}\tanh ^{-1}\left\langle
L\right\rangle _{T_{m}}}{\left[ \left\langle L\right\rangle _{T_{m}}-\left(
1-\left\langle L\right\rangle _{T_{m}}^{2}\right) \tanh ^{-1}\left\langle
L\right\rangle _{T_{m}}\right] ^{2}}\int_{0}^{\left\langle L\right\rangle
_{T_{m}}}\frac{8x^{4}\mathrm{d}x}{\left( 1-x^{2}\right) ^{3}}<0.
\end{equation}%
Therefore, we have the following argument:

\begin{equation*}
\left.
\begin{array}{r}
\left\vert \mathcal{H}_{2}\right\vert \text{ is fixed} \\
T_{m}\nearrow
\end{array}%
\right\} \overset{\text{Eqn.~(\ref{Tm})}}{\Longrightarrow }\left\langle
L\right\rangle _{T_{m}}\searrow
\begin{array}{c}
\\
\left.
\begin{array}{r}
\Longrightarrow \mathcal{H}_{1}-3k_{B}T\Upsilon \nearrow  \\
\left\vert \Upsilon \right\vert \text{ is fixed, }T_{m}\nearrow
\end{array}%
\right\}
\end{array}%
\begin{array}{c}
\\
\Longrightarrow \mathcal{H}_{1}\nearrow .
\end{array}%
\end{equation*}
\end{widetext}Thus, $\left( \partial T_{m}/\partial \mathcal{H}_{1}\right) _{%
\mathcal{H}_{2},\Upsilon }>0$ is evident. $\blacksquare $

That is to say, $T_{m}$ is not only determined by the bonding
energy related to $\mathcal{H}_{2}$, but is also modulated by the
difficulty to create interstitials, which is signified by
$\mathcal{H}_{1}$ -- the energy barrier that hinders information
transfer between atoms. The mathematical inequality above could be
physically interpreted as \textquotedblleft the better information
transfer, the lower melting point\textquotedblright . Unlike the
liquid-gas phase transition involving the thorough dissociation of
bonding atom pairs, in which $J_{2}$ plays a decisive r\^{o}le in
determining the boiling point, the SLPT is possible when
destruction of the lattice structure and the excitation of
interstitials are both highly encouraged. Therefore, some
materials could have high boiling points and \textquotedblleft
disproportionately\textquotedblright\ low melting points,
if they have a large $\left\vert \mathcal{H}_{2}\right\vert $ and a small $%
\left\vert \mathcal{H}_{1}\right\vert $. (This may have shed light on the
peculiar behavior of Ga and In, two metals that melt near room temperature
and boil at thousands of Kelvins, although neither metal falls into the
category of fcc structure.) The lattice softening $\Upsilon >0$ is also
conducive to melting in that $\mathcal{H}_{1}-3k_{B}T\Upsilon <\mathcal{H}%
_{1}$ enhances the apparent $\left\vert \mathcal{H}_{2}\right\vert
/\left\vert \mathcal{H}_{1}\right\vert $ ratio. From this we know that the
atom displacement and lattice softening are mutually complementary and this
rule underlies the intrinsic relation between Lindemann and Born criteria.

The atom-scale pathway (\textit{how}) of melting is spatial
correlations of atom occupancy fluctuations. The
correlation-response scheme proves to be the effective way to
transfer information in the 3D system discussed in this paper. The
anomalous sinusoidal correlation wave results from the
non-Gaussian fluctuations of atoms. In Ref.~\cite{Jin}, it was
observed that melting is preceded by the non-Gaussian behavior of
atom displacement, but the causal relationship between the
deviation from normal distribution and the SLPT was not explicitly
established. This paper presents a clear demonstration of this
causality in Sect.~III.D and E. From the conclusions in those
subsections, we may find the critical diameter of the SDIs for the
example shown in FIG.~\ref{FIG3:c} (b): $\pi \xi _{\min }=15.03a$,
and the
spherical volume $\pi (\pi \xi _{\min })^{3}/6=1780a^{3}$ which contains $%
890 $ atoms in average. This forms a good comparison with the statement in
Ref.~\cite{Jin} that the \textquotedblleft Lindemann
particle\textquotedblright\ clusters consist of $10^{2}-10^{3}$ atoms. The
critical volume of the liquid nucleus $\pi (\pi \xi _{\min })^{3}/6$ will be
evaluated in detail in Appendix B.

By exploring the atom-scale pathway in the thermodynamic limit, we
highlight the fluctuations that propagate as a sinusoidal wave.
This wave cuts the space into separate compartments exhibiting
ostentatiously mutual independence, and is thus responsible for
the illusion that the inhomogeneous instability suddenly emerges
from nowhere. There is no contradiction between the nucleation and
the thermodynamic limit, after all. This answers Question 2.

The melting point (\textit{when}) is defined in Eqn.~(\ref{Tm}) and it is
shown to be the superheating limit predicted by Born and Lindemann criteria
simultaneously. When rewritten, the melting formula takes the form
\begin{equation}
T_{m}=\frac{3\left| J_{2}\right| \left\langle L\right\rangle _{T_{m}}^{3}}{%
2k_{B}\left( \frac{\left\langle L\right\rangle _{T_{m}}}{1-\left\langle
L\right\rangle _{T_{m}}^{2}}-\tanh ^{-1}\left\langle L\right\rangle
_{T_{m}}\right) }<\frac{9\left| J_{2}\right| }{4k_{B}}.
\end{equation}%
The inequality results from the following

\noindent \textit{Theorem:}
\begin{equation}
\frac{2}{3}<\frac{\frac{\left\langle L\right\rangle _{T_{m}}}{1-\left\langle
L\right\rangle _{T_{m}}^{2}}-\tanh ^{-1}\left\langle L\right\rangle _{T_{m}}%
}{\left\langle L\right\rangle _{T_{m}}^{3}}
\end{equation}%
\begin{widetext}
\noindent \textit{Proof:} Pick $\left\langle L\right\rangle _{T_{m}}$ $>0,$%
\begin{eqnarray}
&&\frac{\left\langle L\right\rangle _{T_{m}}}{1-\left\langle L\right\rangle
_{T_{m}}^{2}}-\tanh ^{-1}\left\langle L\right\rangle _{T_{m}}-\frac{2}{3}%
\left\langle L\right\rangle _{T_{m}}^{3}=\int_{0}^{\left\langle
L\right\rangle _{T_{m}}}\frac{2x^{4}\left( 2-x^{2}\right)
\mathrm{d}x}{\left( 1-x^{2}\right) ^{2}}>0
\end{eqnarray}%
$\blacksquare $

\smallskip We can check the reasonability of the formula above by the
following table:

\begin{center}
\begin{tabular}{|cc|ccccc|ccc|}
\hline
&  & Ne & Ar & Kr & Xe & Rn & Cu & Ag & Au \\ \hline\hline
\multicolumn{1}{|l}{$T_{E}$} & \multicolumn{1}{l|}{$/\mathrm{K}$} &
\multicolumn{1}{|r}{24.56} & \multicolumn{1}{r}{83.8} & \multicolumn{1}{r}{
115.79} & \multicolumn{1}{r}{161.4} & \multicolumn{1}{r|}{202} &
\multicolumn{1}{|r}{1357.6} & \multicolumn{1}{r}{1234.93} &
\multicolumn{1}{r|}{1337.33} \\
\multicolumn{1}{|l}{$T_{\text{b}}$} & \multicolumn{1}{l|}{$/\mathrm{K}$} &
\multicolumn{1}{|r}{27.07} & \multicolumn{1}{r}{87.3} & \multicolumn{1}{r}{
119.93} & \multicolumn{1}{r}{165.1} & \multicolumn{1}{r|}{211.3} &
\multicolumn{1}{|r}{2840} & \multicolumn{1}{r}{2435} & \multicolumn{1}{r|}{
3129} \\
\multicolumn{1}{|l}{$\Delta H_{\text{vap}}$} & \multicolumn{1}{l|}{$/\mathrm{%
kJ\ mol}^{-1}$} & \multicolumn{1}{|r}{1.7326} & \multicolumn{1}{r}{6.447} &
\multicolumn{1}{r}{9.029} & \multicolumn{1}{r}{12.636} & \multicolumn{1}{r|}{
16.4} & \multicolumn{1}{|r}{300.3} & \multicolumn{1}{r}{250.58} &
\multicolumn{1}{r|}{334.4} \\
\multicolumn{1}{|l}{$\Delta H_{\text{fus}}$} & \multicolumn{1}{l|}{$/\mathrm{%
kJ\ mol}^{-1}$} & \multicolumn{1}{|r}{0.3317} & \multicolumn{1}{r}{1.188} &
\multicolumn{1}{r}{1.638} & \multicolumn{1}{r}{2.297} & \multicolumn{1}{r|}{
2.89} & \multicolumn{1}{|r}{13.05} & \multicolumn{1}{r}{11.3} &
\multicolumn{1}{r|}{12.55} \\
\multicolumn{1}{|l}{$\Delta S_{\text{fus}}$} & \multicolumn{1}{l|}{$/\mathrm{%
J\ mol}^{-1}\mathrm{K}^{-1}$} & \multicolumn{1}{|r}{13.5} &
\multicolumn{1}{r}{14.2} & \multicolumn{1}{r}{14.1} & \multicolumn{1}{r}{14.2
} & \multicolumn{1}{r|}{14.3} & \multicolumn{1}{|r}{9.61} &
\multicolumn{1}{r}{9.15} & \multicolumn{1}{r|}{9.38} \\
\multicolumn{1}{|l}{$\left\vert J_{2}\right\vert N_{A}\mathrm{\ }$} &
\multicolumn{1}{l|}{$/\mathrm{kJ\ mol}^{-1}$} & \multicolumn{1}{|r}{0.3441}
& \multicolumn{1}{r}{1.273} & \multicolumn{1}{r}{1.778} & \multicolumn{1}{r}{
2.489} & \multicolumn{1}{r|}{3.215} & \multicolumn{1}{|r}{52.23} &
\multicolumn{1}{r}{43.65} & \multicolumn{1}{r|}{57.83} \\
\multicolumn{1}{|l}{$\frac{9\left\vert J_{2}\right\vert }{4k_{B}}$} &
\multicolumn{1}{l|}{$/\mathrm{K}$} & \multicolumn{1}{|r}{93.109} &
\multicolumn{1}{r}{344.37} & \multicolumn{1}{r}{481.13} & \multicolumn{1}{r}{
673.58} & \multicolumn{1}{r|}{870.07} & \multicolumn{1}{|r}{14134} &
\multicolumn{1}{r}{11812} & \multicolumn{1}{r|}{15649} \\
\multicolumn{1}{|l}{$T_{m}$} & \multicolumn{1}{l|}{$/\mathrm{K}$} &
\multicolumn{1}{|r}{29.47} & \multicolumn{1}{r}{100.6} & \multicolumn{1}{r}{
138.9} & \multicolumn{1}{r}{193.7} & \multicolumn{1}{r|}{242.4} &
\multicolumn{1}{|r}{1629} & \multicolumn{1}{r}{1482} & \multicolumn{1}{r|}{
1604} \\
\multicolumn{2}{|l|}{$\frac{1}{2}(1-\left\vert \left\langle L\right\rangle
_{T_{m}}\right\vert )$} & \multicolumn{1}{|r}{0.0475} & \multicolumn{1}{r}{
0.0440} & \multicolumn{1}{r}{0.0435} & \multicolumn{1}{r}{0.0430} &
\multicolumn{1}{r|}{0.0420} & \multicolumn{1}{|r}{0.0180} &
\multicolumn{1}{r}{0.0195} & \multicolumn{1}{r|}{0.0160} \\
\multicolumn{1}{|l}{$\Delta S_{m}$} & \multicolumn{1}{r|}{$/\mathrm{J\ mol}%
^{-1}\mathrm{K}^{-1}$} & \multicolumn{1}{|r}{8.35} & \multicolumn{1}{r}{8.53}
& \multicolumn{1}{r}{8.56} & \multicolumn{1}{r}{8.57} & \multicolumn{1}{r|}{
8.63} & \multicolumn{1}{|r}{10.0} & \multicolumn{1}{r}{9.93} &
\multicolumn{1}{r|}{10.2} \\ \hline
\end{tabular}
\end{center}

In this table, all the elements (Ne, Ar, Kr, Xe and Rn from the 0
group; Cu, Ag and Au from the IB group) assume fcc structures in
the solid phase. The first four rows were taken from the
descriptions of individual elements in the online encyclopedia,
http://en.wikipedia.org/ (accessed Sept. 10, 2004). $T_{E}$ is the
equilibrium melting point of
crystals with surfaces. $T_{\text{b}}$ is the boiling point of the liquid. $%
\Delta H_{\text{vap}}$ is the heat of evaporation. $\Delta
H_{\text{fus}}$ is the heat of fusion. The data in the other rows
are obtained/estimated as follows: $\Delta S_{\text{fus}}=\Delta
H_{\text{fus}}/T_{E}$ is the entropy change due to fusion;
$T_{m}=1.2T_{E}$ is an estimate of superheating limit based
on Ref.~\cite{Jin}; $\left\vert J_{2}\right\vert =(\Delta H_{\text{vap}%
}+\Delta H_{\text{fus}})/6N_{A}$ is a rough estimate for the $J_2$
parameter, where $N_{A}$ is the Avogadro's number;
$\frac{1}{2}(1-\left\vert \left\langle L\right\rangle
_{T_{m}}\right\vert )$ is the critical concentration of
interstitial defects that initiates melting, where $\left\vert
\left\langle L\right\rangle
_{T_{m}}\right\vert $ is calculated from $\left\vert J_{2}\right\vert $, $%
T_{m}$ and Eqn.~(\ref{Tm});

\begin{eqnarray}
\Delta S_{m} &=&\left. S^{\text{conf}}\right\vert _{L\left( \mathbf{r}%
\right) \equiv 0}-\left. S^{\text{conf}}\right\vert _{L\left( \mathbf{r}%
\right) \equiv \left\vert \left\langle L\right\rangle _{T_{m}}\right\vert }
\notag \\
&=&N_{A}k_{B}\left[ 2\log 2+\left( 1-\left\vert \left\langle L\right\rangle
_{T_{m}}\right\vert \right) \log \frac{1-\left\vert \left\langle
L\right\rangle _{T_{m}}\right\vert }{2}+\left( 1+\left\vert \left\langle
L\right\rangle _{T_{m}}\right\vert \right) \log \frac{1+\left\vert
\left\langle L\right\rangle _{T_{m}}\right\vert }{2}\right]
\end{eqnarray}
In principle,
\begin{equation}
\Delta S_{\text{fus}}=\left. S^{\text{conf}}\right\vert _{L\left( \mathbf{r}%
\right) \equiv 0}-\left. S^{\text{conf}}\right\vert _{L\left( \mathbf{r}%
\right) \equiv \left\vert \left\langle L\right\rangle _{T_{E}}\right\vert
}+N_{A}k_{B}\log \frac{V^{l}}{V^{s}},
\end{equation}%
\end{widetext}where $V^{l}$ and $V^{s}$ are the molar volume of liquid and
solid, respectively. Therefore, the disagreement between $\Delta S_{\text{fus%
}}$ and $\Delta S_{m}$ may be explained as the volume change and the $%
T_{m}/T_{E} $ ratio that are not fairly covered in the above calculation of $%
\Delta S_{m} $. In general, the table justifies the inequality $%
T_{m}<9\left\vert J_{2}\right\vert /4k_{B}$, and shows similar $\Delta S_{%
\text{fus}}$ (\textit{or }$\Delta S_{m}$) values for the elements in the
same group. The data also suggest that melting is triggered when the
concentration of interstitial defects is very low but still high enough \cite%
{hconc} to induce correlation and cooperation.

For noble gases Ne, Ar, Kr, Xe, Rn, the interaction mode is the well-known
van der Waals force and is described by the Lennard-Jones function \cite{Jin}%
, so they share theoretically the same (and practically similar)
dimensionless parameters $\left\vert \mathcal{H}_{2}\right\vert
/\left\vert \mathcal{H}_{1}\right\vert $, $\Upsilon $ and
$V^{l}/V^{s}$ in our model. Hence they should have the same
dimensionless values: $\left\vert \left\langle L\right\rangle
_{T_{m}}\right\vert $, $\left\vert J_{2}\right\vert /k_{B}T_{m}\
$and $\Delta S_{m}/N_{A}k_{B}$. Another dimensionless parameter
$\delta _{L}^{\ast }$ could be obtained by using the following

\noindent \textit{Theorem:}
\begin{equation}
3k_{B}T_{m}=4\pi ^{2}\langle \nu ^{2}\rangle m(a\delta _{L}^{\ast })^{2}
\end{equation}
where $\langle \nu ^{2}\rangle $ is the average frequency \cite{Lindemann}$.$

\noindent \textit{Proof:} By multiplying the Newtonian equation $m\mathbf{%
\ddot{r}}(t)=\mathbf{F}\left( t\right) $ with $\mathbf{r}(t)$ and taking
time average, we can establish the following Langevin equation
\begin{equation}
m\overline{\mathbf{r}(t)\cdot \mathbf{\ddot{r}}(t)}=\overline{\mathbf{r}%
(t)\cdot \mathbf{F}\left( t\right) },
\end{equation}%
where $m$ is the mass of the atom, on which a force $\mathbf{F}\left(
t\right) $ is exerted at time $t$.
\begin{eqnarray}
\overline{\mathbf{r}(t)\cdot \mathbf{\ddot{r}}(t)} &=&\lim_{\tau \rightarrow
\infty }\frac{1}{\tau }\int_{0}^{\tau }\mathbf{r}(t)\cdot \mathbf{\ddot{r}}%
(t)\mathrm{d}t  \notag \\
&=&\lim_{\tau \rightarrow \infty }\frac{1}{\tau }\int_{0}^{\tau }\mathbf{r}%
(t)\cdot \mathrm{d}\mathbf{\dot{r}}(t)  \notag \\
&=&\lim_{\tau \rightarrow \infty }\frac{1}{\tau }\left[ \left. \mathbf{r}%
(t)\cdot \mathbf{\dot{r}}(t)\right\vert _{t=0}^{\tau }\right] -\lim_{\tau
\rightarrow \infty }\frac{1}{\tau }\int_{0}^{\tau }\mathbf{\dot{r}}^{2}(t)%
\mathrm{d}t  \notag \\
&&
\end{eqnarray}%
$\left. \mathbf{r}(t)\cdot \mathbf{\dot{r}}(t)\right\vert _{t=0}^{\tau }$ is
finite insofar as vibration and Brownian motion are concerned. So $%
\lim_{\tau \rightarrow \infty }(1/\tau )\left. \mathbf{r}(t)\cdot \mathbf{%
\dot{r}}(t)\right\vert _{t=0}^{\tau }=0$, and $\overline{\mathbf{r}(t)\cdot
\mathbf{\ddot{r}}(t)}=-\overline{\mathbf{\dot{r}}^{2}(t)}$. According to the
ergodicity argument, we may replace the time average by the ensemble average
as long as $T<T_{m}$. Therefore,

\begin{eqnarray}
\bigskip \underset{||}{m\overline{\mathbf{r}(t)\cdot \mathbf{\ddot{r}}(t)}}
&=&-m\overline{\mathbf{\dot{r}}^{2}(t)}=-m\left\langle \mathbf{\dot{r}}%
^{2}\right\rangle =-3k_{B}T  \notag \\
\overline{\mathbf{r}(t)\cdot \mathbf{F}\left( t\right) } &=&\left\langle
\mathbf{r}\cdot \mathbf{F}\right\rangle =-4\pi ^{2}m\left\langle \nu ^{2}%
\mathbf{r}\cdot \mathbf{r}\right\rangle
\end{eqnarray}%
Define $\langle \nu ^{2}\rangle =\left\langle \nu ^{2}\mathbf{r}%
^{2}\right\rangle /\left\langle \mathbf{r}^{2}\right\rangle $ \cite%
{Lindemann}, then we have%
\begin{eqnarray}
\underset{||}{\lim_{T\rightarrow T_{m}-0}3k_{B}T} &=&\underset{||}{%
\lim_{T\rightarrow T_{m}-0}4\pi ^{2}m\left\langle \nu ^{2}\right\rangle
\left\langle \mathbf{r}^{2}\right\rangle }  \notag \\
3k_{B}T_{m} &=&4\pi ^{2}m\left\langle \nu ^{2}\right\rangle \left( a\delta
_{L}^{\ast }\right) ^{2}
\end{eqnarray}%
$\blacksquare $

For Lennard-Jones potential, $m\left\langle \nu ^{2}\right\rangle \propto
\left\vert \mathcal{H}_{2}\right\vert \propto T_{m}$, so there is a
universal $\delta _{L}^{\ast }$ for all fcc crystals governed by van der
Waals forces. In our model, $k_{B}T_{m}/\left\vert J_{2}\right\vert $ is a
function of $\left\vert \left\langle L\right\rangle _{T_{m}}\right\vert $
that varies slowly with respect to $\left\vert \left\langle L\right\rangle
_{T_{m}}\right\vert $. The variation itself infers that $\delta _{L}^{\ast }$
is not a universal constant independent of interaction force, and the slow
variation explains the reason why similar materials have nearly identical $%
\delta _{L}^{\ast }$'s. In a nutshell, the most influential factors for the
critical Lindemann ratio $\delta _{L}^{\ast }$ are the profile (\emph{not}
every detail!) of the interatomic force that determines $\langle L\rangle
_{T_{m}}$ as a function of $J_{1},J_{2}$ and $\Upsilon $ in different
crystals. There is no such a strictly universal $\delta _{L}^{\ast }$ for
fcc lattice \cite{Lrule}. This answers Question 3.

We admit that the current model solution may need further refinement in
terms of parameter estimate. The relations $\mathcal{H}_{2}=(3/8)J_{2}$, $%
\gamma =a^{2}\mathcal{H}_{1}/2$ and the representation of $\pi \xi _{\min }$
in this work are obtained by bold assumptions and simplifications. However,
judging the physical origin of the collaborative effect and domain wall
energy, it is clear that $\mathcal{H}_{2}<0$, $\gamma >0$ is not
questionable even when they are evaluated exactly. $\rho _{3}>0$ is also
guaranteed in all cases:

\noindent \textit{Theorem:} $\rho _{3}>0$ at $T=T_{m}$

\noindent \textit{Proof:}
\begin{eqnarray}
6\rho _{3} &=&\frac{\partial ^{3}}{\partial ^{3}\lambda }\left( 2\mathcal{H}%
_{2}\lambda ^{3}+k_{B}T_{m}\tanh ^{-1}\lambda \right)  \notag \\
&=&12\mathcal{H}_{2}+2k_{B}T_{m}\frac{4\lambda ^{2}+\left( 1-\lambda
^{2}\right) }{\left( 1-\lambda ^{2}\right) ^{3}}  \notag \\
&=&k_{B}T_{m}\left[ -\frac{3}{\lambda ^{3}}\left( \frac{\lambda }{1-\lambda
^{2}}-\tanh ^{-1}\lambda \right) +2\frac{3\lambda ^{2}+1}{\left( 1-\lambda
^{2}\right) ^{3}}\right]  \notag \\
&=&\frac{48k_{B}T_{m}}{\lambda ^{3}}\int_{0}^{\lambda }\frac{x^{4}\mathrm{d}x%
}{\left( 1-x^{2}\right) ^{4}}>0,
\end{eqnarray}%
where $\lambda =\left\langle L\right\rangle _{T_{m}}$. $\blacksquare $

This immediately justitfies the representation of $\pi \xi \sim \sqrt{2}a%
\mathcal{H}_{1}^{2}(2\rho _{3}k_{B}T_{m})^{-1}>0$. Both the diagram in FIG.~%
\ref{FIG4:d}(b) and analytical expression here show that%
\begin{equation}
\lim_{\left\langle L\right\rangle _{T_{m}}\rightarrow 0}\frac{\rho _{3}}{%
k_{B}T_{m}}=0.
\end{equation}%
Hence,
\begin{equation}
\lim_{\left\langle L\right\rangle _{T_{m}}\rightarrow 0}\pi \xi =\infty .
\end{equation}%
This reminds us of the infinite correlation length that characterizes the
critical point of a \emph{continuous} phase transition -- now a limit case
of a sequence of first-order SLPTs.

Fortunately, the effectiveness of most of the inferences in our
model depends on the sign of the parameters instead of their exact
values, so the related physical picture of melting will not change
even if parameter estimates are modified by better methods.

\section{Conclusions}

In Paper II, we have provided the detailed arguments involved in the SLPT
model for a 3D fcc elemental crystal. We have used two independent
approaches to obtain the same equation for the evolution of defect
concentration. The solution of the model shows that the phenomenological
concerns in Lindemann and Born criteria: atom displacement and lattice
softening alone are only conducive to, but not decisive in the melting
process. The two criteria bespeak the underlying cooperative creation of
interstitials and inter-defect correlations, which are crucial for the
feasibility of a first-order SLPT marked by catastrophes of both rigidity
and displacement. The sinusoidal correlation waves give rise to interstitial
clusters near the melting point and trigger the heterogeneous formation of
the SDIs -- the pre-liquid droplets satisfying Born and Lindemann criteria
simultaneously.

\appendix

\section{The Defect Concentration Equation Revisited}

\subsection{``Chemical Equilibrium'' Without $\mathcal{H}%
_{2}$ and $\Upsilon $}

In this subsection, we will provide an alternative derivation of the
``chemical equilibrium'' condition Eqn.~(\ref{chem_eq}) so as to double-check Eqn.~(%
\ref{eqmotion}). For simplicity, vibration is provisionally neglected in the
following few paragraphs.

We set out to evaluate the configurational energy difference
between defective and non-defective cells when $(1+\left|
\left\langle L\right\rangle _{T}\right| )/2=P$ and $(1-\left|
\left\langle L\right\rangle _{T}\right| )/2=q$. Here, $P$ is the
probability that a site is occupied ``correctly'' and $q=1-P$. We
use the following diagrams to aid calculation, bearing in mind
that each edge in the cube below is shared by four cubes and each
diagonal line on the surface of the cube is shared by two cubes.
Every vertex of the cube (shared by eight cubes) is labeled by the
site occupancy -- the probability that the corresponding site is
occupied by an atom.

If NNN correlation is negligible, we may use the following graph
counting to evaluate the average energy difference between a
non-defective and a defective cell:
\begin{equation}
-\frac{\Delta \overline{\varepsilon }}{4}=\frac{1}{4}\left( \overline{%
\varepsilon }_{\text{non-defective cell}}-\overline{\varepsilon }_{\text{%
defective cell}}\right)
\end{equation}%
\begin{widetext}
\begin{eqnarray*}
 &=& \bpm
\includegraphics[width=2cm]{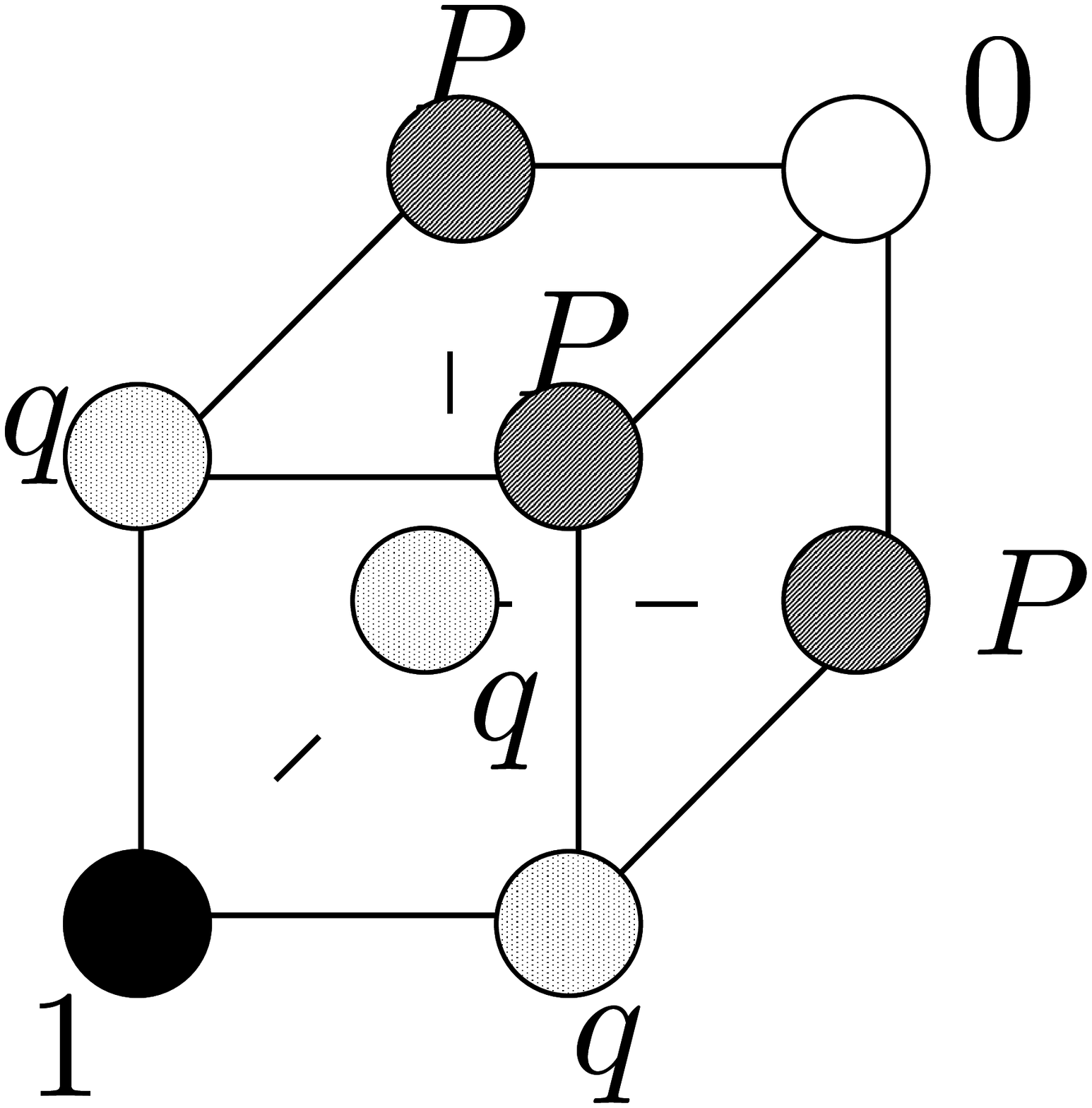} \epm\  - \bpm
\includegraphics[width=2cm]{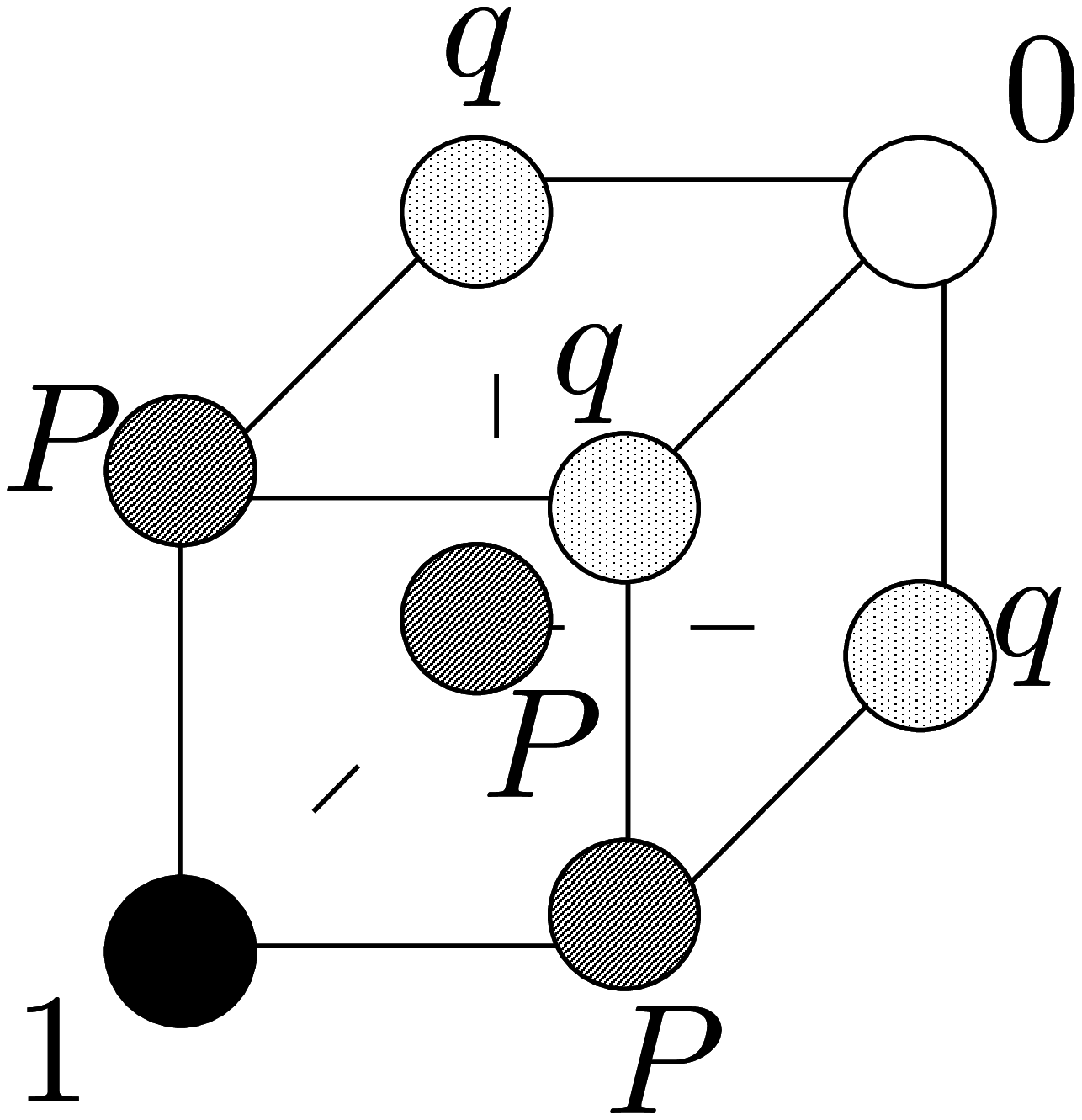} \epm\ \notag \\
&=&\left[3\bpm \includegraphics[width=2cm]{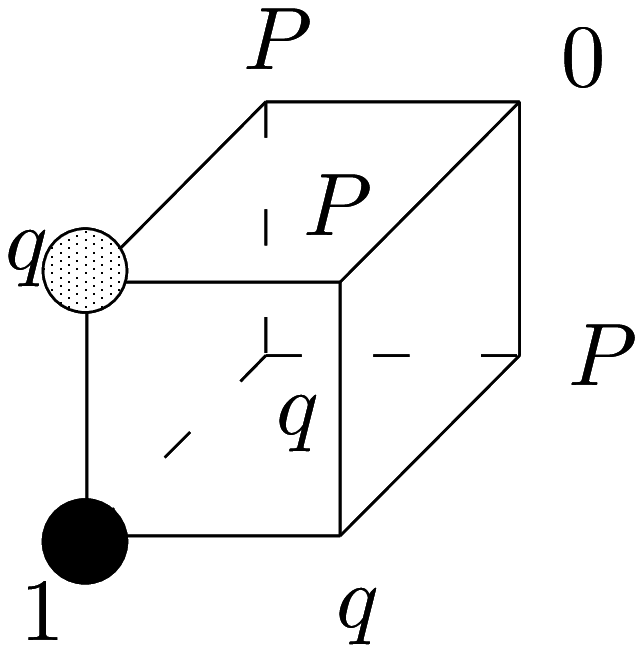} \epm +3\bpm
\includegraphics[width=2cm]{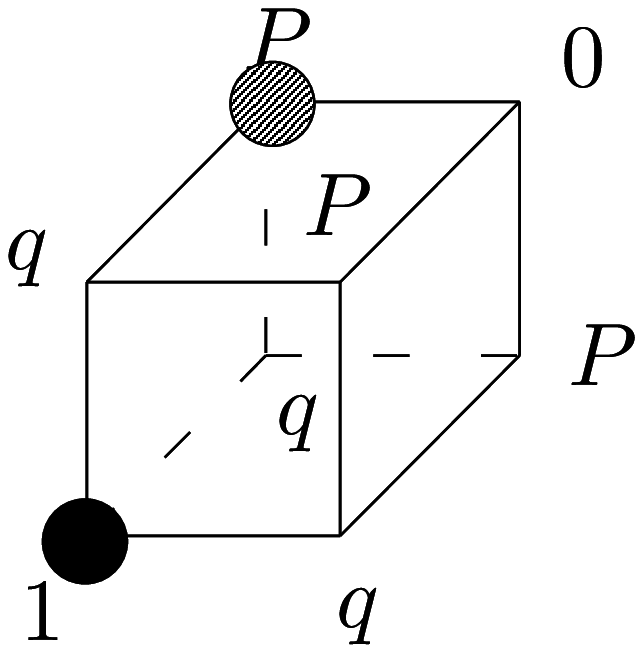} \epm+6\bpm
\includegraphics[width=2cm]{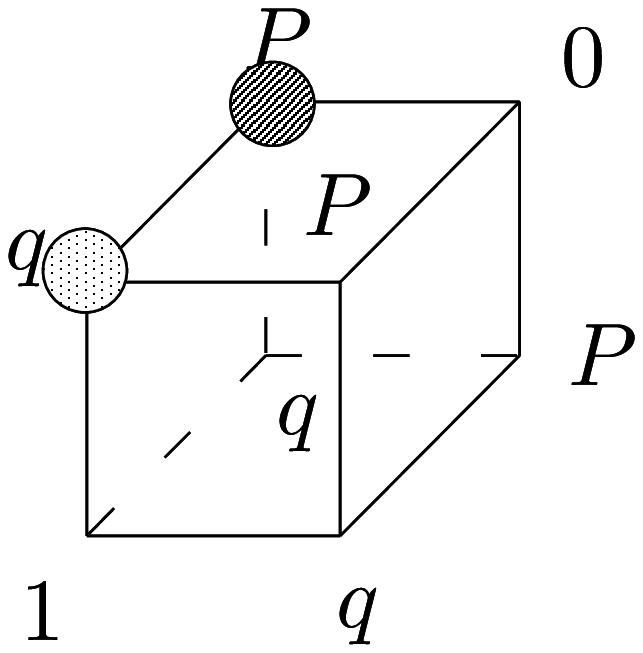} \epm\
+3\bpm \includegraphics[width=2cm]{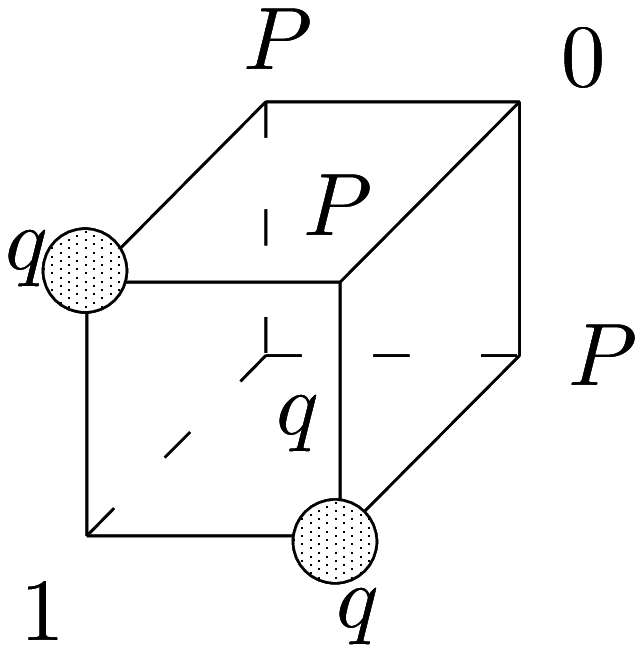} \epm+3\bpm
\includegraphics[width=2cm]{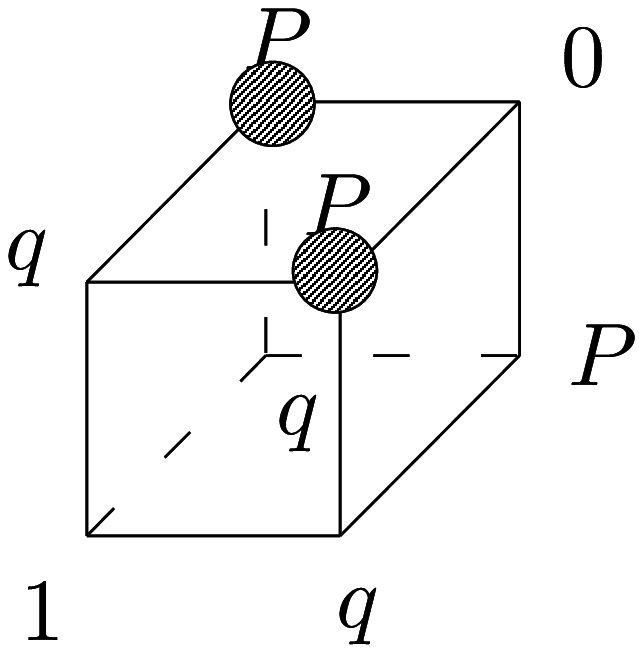} \epm\right] \notag \\
&&-\left[3\bpm
\includegraphics[width=2cm]{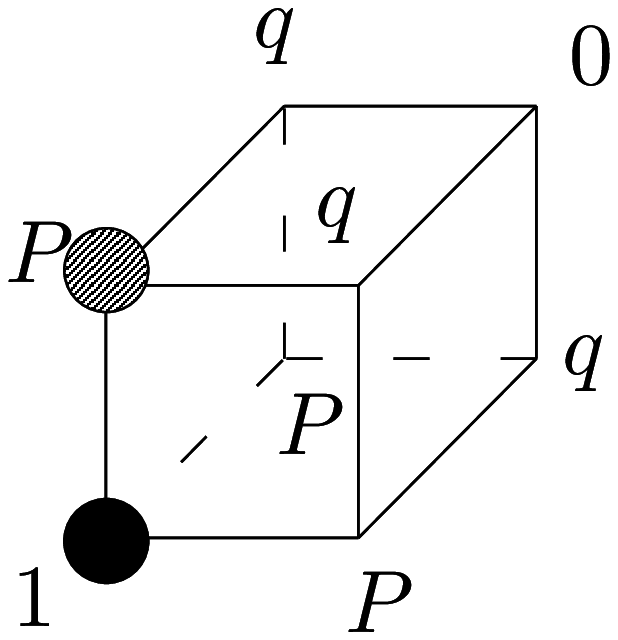} \epm+3\bpm \includegraphics[width=2cm]{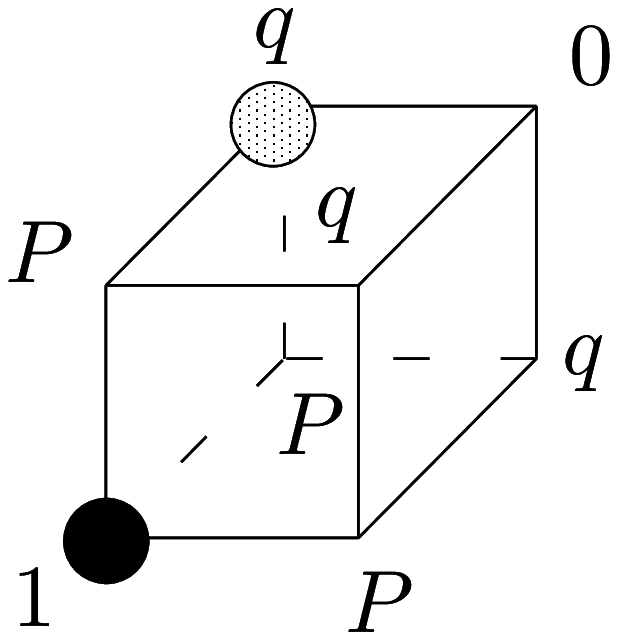} \epm +6
\bpm \includegraphics[width=2cm]{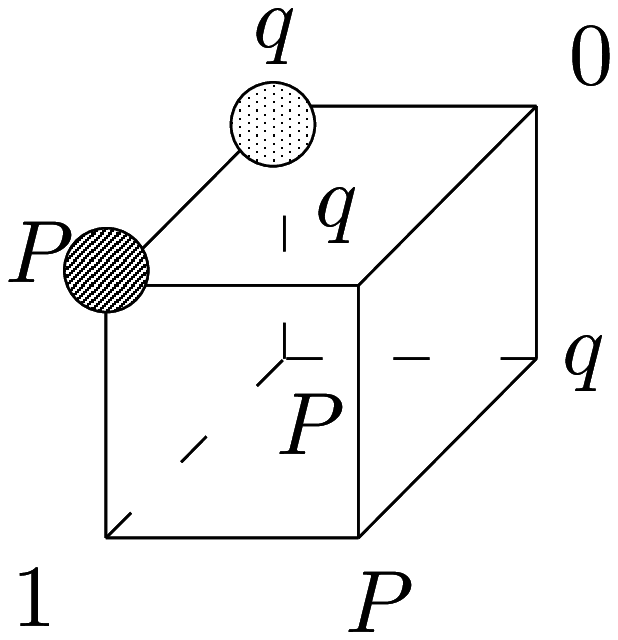} \epm+3\bpm
\includegraphics[width=2cm]{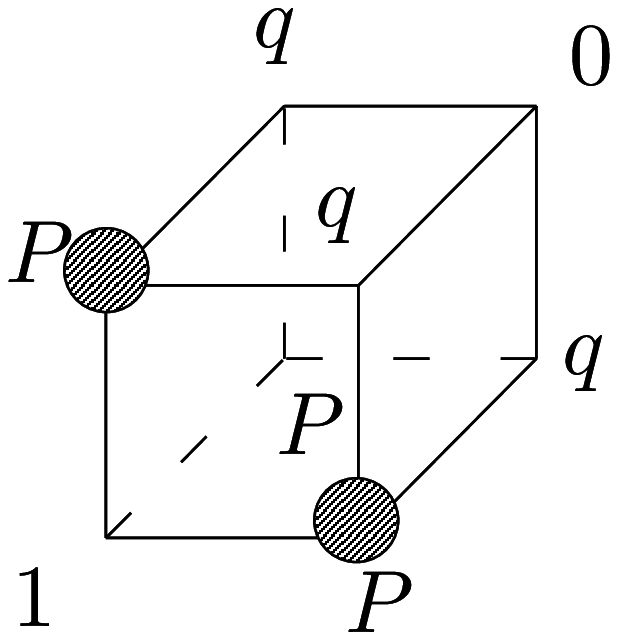} \epm+3\bpm
\includegraphics[width=2cm]{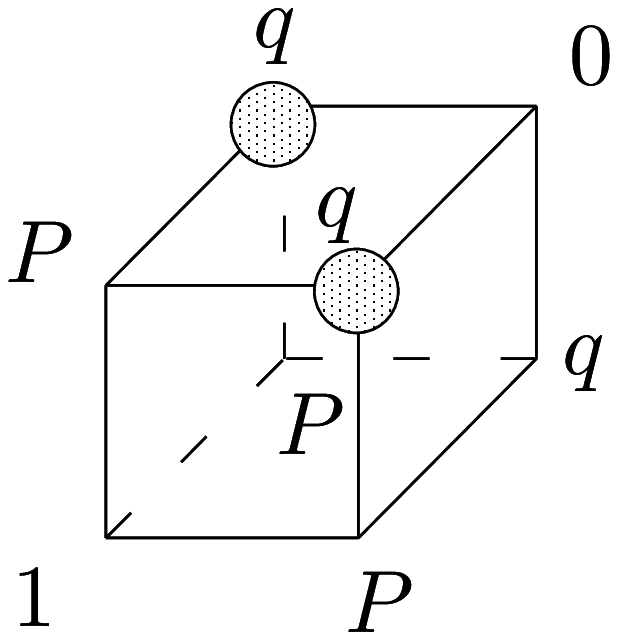} \epm\right]
\end{eqnarray*}
\end{widetext}%
\begin{eqnarray}
&=&\left(3\frac{J_{1}}{4}q+3\frac{J_{2}}{2}P+6\frac{J_{1}}{4}qP+3\frac{J_{2}%
}{2}q^{2}+3\frac{J_{2}}{2}P^{2}\right)  \notag \\
&&-\left(3\frac{J_{1}}{4}P+3\frac{J_{2}}{2}q+6\frac{J_{1}}{4}Pq+3\frac{J_{2}%
}{2}P^{2}+3\frac{J_{2}}{2}q^{2}\right)  \notag \\
&=&\frac{3}{4}\left( -J_{1}+2J_{2}\right) \left( P-q\right)  \notag \\
&=&\frac{3}{4}\left( -J_{1}+2J_{2}\right) \left\vert \left\langle
L\right\rangle _{T}\right\vert .
\end{eqnarray}%
This value is equal to $-(1/4)\Delta \overline{\varepsilon }$ because for
each cube, the state of $2\times (1/8)=1/4$ atoms are definite, marked by
occupancy 1 and 0 respectively. Therefore, the dynamic equilibrium between
the two types of cell requires that
\begin{equation}
\frac{q}{P}=\frac{1-\left\vert \left\langle L\right\rangle _{T}\right\vert }{%
1+\left\vert \left\langle L\right\rangle _{T}\right\vert }=\exp \left( -%
\frac{2\mathcal{H}_{1}\left\vert \left\langle L\right\rangle _{T}\right\vert
}{k_{B}T}\right)  \label{qp}
\end{equation}%
where $\mathcal{H}_{1}=\left( 3/2\right) \left( J_{1}-2J_{2}\right) $. This
agrees with Eqs. (\ref{eqmotion}) and (\ref{chem_eq}) when terms associated
with $\mathcal{H}_{2}$, $\Upsilon $ are neglected. As $\left\vert
\left\langle L\right\rangle _{T}\right\vert \rightarrow 1$, Eqn.~(\ref{qp})
implies $q\sim \exp \left( -u/2k_{B}T\right) $, which agrees with the
well-established theory of Frenkel defects. Here, $u=4\mathcal{H}_{1}$ is
the excitation energy of an interstitial defect.

\subsection{``Chemical Equilibrium'' involving $\mathcal{H}_{2}$ and $%
\Upsilon $}

When $\left| \left\langle L\right\rangle _{T}\right| $ is
sufficiently far from unity, the cooperation and correlation
between NNN atoms could not be neglected. We can estimate such NNN
correlation by postulating that the three sites that are NNN
neighbor to the probability-one-occupied site tend to be occupied
or be unoccupied simultaneously. Considering this cooperation
effect, we should modify $-(1/4)\Delta \overline{\varepsilon }$ by
adding the contribution from the following diagrams:

\begin{widetext}
\begin{equation*}
\left[\bpm
\includegraphics[width=2cm]{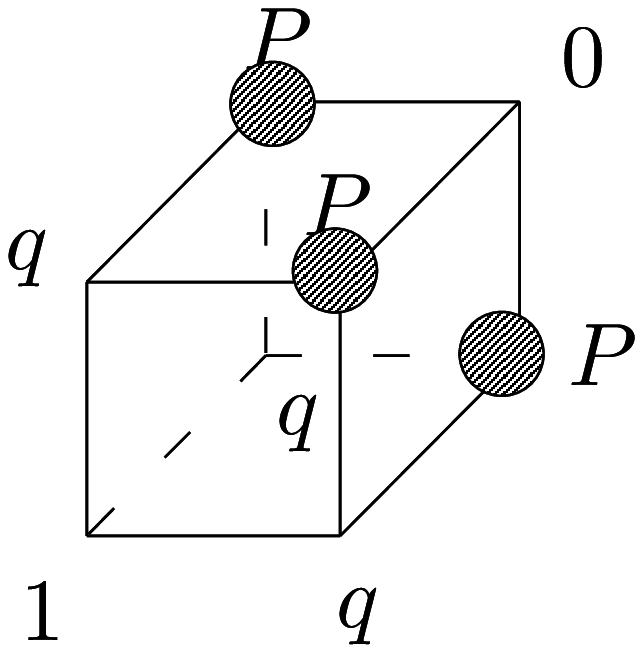} \epm-3\bpm
\includegraphics[width=2cm]{CHEQ07.EPS} \epm\right]-\left[\bpm
\includegraphics[width=2cm]{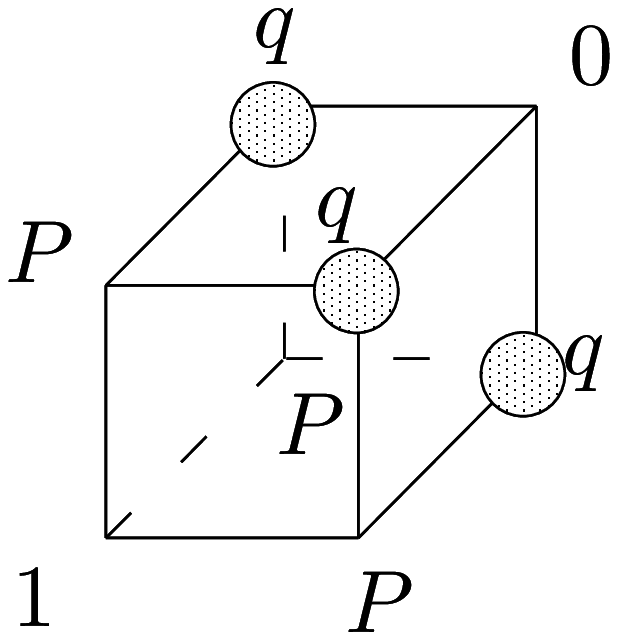} \epm-3 \bpm
\includegraphics[width=2cm]{CHEQ12.EPS} \epm\right]
\end{equation*}
\end{widetext}%
\begin{eqnarray}
&=&\left( \frac{3J_{2}}{2}P^{3}-3\frac{J_{2}}{2}P^{2}\right) -\left( \frac{%
3J_{2}}{2}q^{3}-3\frac{J_{2}}{2}q^{2}\right)  \notag \\
&=&-\frac{3J_{2}}{2}\left( P-q\right) Pq  \notag \\
&=&\frac{3J_{2}}{8}\left\vert \left\langle L\right\rangle _{T}\right\vert
(\left\vert \left\langle L\right\rangle _{T}\right\vert ^{2}-1).
\end{eqnarray}%
Accordingly, the chemical equilibrium condition now reads:
\begin{equation}
\frac{1-\left\vert \left\langle L\right\rangle _{T}\right\vert }{%
1+\left\vert \left\langle L\right\rangle _{T}\right\vert }=e^{-\frac{1}{%
k_{B}T}[ 2\mathcal{H}_{1}| \left\langle L\right\rangle _{T}|
-4\mathcal{H}_{2}| \left\langle L\right\rangle _{T}| (|
\left\langle L\right\rangle _{T}| ^{2}-1)] }
\end{equation}%
where $\mathcal{H}_{2}=\left( 3/8\right) J_{2}$, in perfect agreement with
Eqs. (\ref{eqmotion}) and (\ref{chem_eq}) leaving alone the vibration term.

The contribution from vibrations could be easily considered by the
transformation $4\mathcal{H}_1\mapsto 4\mathcal{H}_1-3k_BT\log (\left\langle
\nu _l\right\rangle /\left\langle \nu _i\right\rangle )$ because averagely
speaking, exciting one Frenkel pair changes one eigenfrequency from $%
\left\langle \nu _l\right\rangle $ to $\left\langle \nu _i\right\rangle $.

The chemical equilibrium between defective and non-defective cells
is reached by stochastic motion of interstitials, which guarantees
sufficient randomness of NN pair states (a corollary of short
relaxation time of NN atom pairs) thereby supporting the former
argument that holds NN correlation to be negligible. In the mean
time, NNN atom pairs have considerably long relaxation time and
their correlation will lead to spatial inhomogeneity, which is to
be elaborated in the Appendix~B.

\section{The Spatial Correlation of Excitations in Detail}

\subsection{The Green Function}

In this subsection, we establish the equation of order correlation
functions and obtain expressions for the correlation length at low
temperatures. We now start to work with the unit system where
$a=\sqrt{2}$. In this special system, $\int
\mathrm{d}^{3}\mathbf{r}$ could be interpreted as the summation
over the $2N$ sites as well as the volume integration.

The equation of $L\left( \mathbf{r}\right) $ spatial correlation Green
function is obtained by the following procedure \cite{Nigel Goldenfeld}:

(1) We introduce a phenomenological external field $H\left( \mathbf{r}%
\right) $ and rewrite the free energy functional as
\begin{equation}
F\left[ L\left( \mathbf{r}\right) ,H\left( \mathbf{r}\right) \right] =F\left[
L\left( \mathbf{r}\right) \right] -\int \mathrm{d}^{3}\mathbf{r\ }H\left(
\mathbf{r}\right) L\left( \mathbf{r}\right) .
\end{equation}

(2) We use functional derivative $\delta $ to verify
\begin{eqnarray}
&&-\frac{\delta ^{2}F\left[ L\left( r\right) ,H\left( \mathbf{r}\right) %
\right] }{\delta H\left( \mathbf{r}\right) \delta H\left( \mathbf{r}^{\prime
}\right) }  \notag \\
&=&k_{B}T\left[ \frac{1}{Q}\frac{\delta ^{2}Q}{\delta H\left( \mathbf{r}%
\right) \delta H\left( \mathbf{r}^{\prime }\right) }-\frac{1}{Q}\frac{\delta
Q}{\delta H\left( \mathbf{r}\right) }\frac{1}{Q}\frac{\delta Q}{\delta
H\left( \mathbf{r}^{\prime }\right) }\right]  \notag \\
&=&\frac{1}{k_{B}T}\left[ \left\langle L\left( \mathbf{r}\right) L\left(
\mathbf{r}^{\prime }\right) \right\rangle -\left\langle L\left( \mathbf{r}%
\right) \right\rangle \left\langle L\left( \mathbf{r}^{\prime }\right)
\right\rangle \right]  \notag \\
&=&\frac{1}{k_{B}T}G\left( \mathbf{r,r}^{\prime }\right)
\end{eqnarray}%
where $Q=\exp \left( -F/k_{B}T\right) $.

(3) On the other hand, $\delta F\left[ L\left( r\right) ,H\left( \mathbf{r}%
\right) \right] /\delta L\left( \mathbf{r}\right) =0$ implies that
\begin{equation}
-\gamma \nabla ^{2}L\left( \mathbf{r}\right) -\rho \left( L\left( \mathbf{r}%
\right) \right) -H\left( \mathbf{r}\right) =0.
\end{equation}%
Recalling that
\begin{equation}
\frac{\delta \left\langle L\left( \mathbf{r}\right) \right\rangle }{\delta
H\left( \mathbf{r}^{\prime }\right) }=-\frac{\delta ^{2}F\left[ L\left(
r\right) ,H\left( \mathbf{r}\right) \right] }{\delta H\left( \mathbf{r}%
\right) \delta H\left( \mathbf{r}^{\prime }\right) }
\end{equation}%
and
\begin{equation}
\frac{\delta H\left( \mathbf{r}\right) }{\delta H\left( \mathbf{r}^{\prime
}\right) }=\delta \left( \mathbf{r}-\mathbf{r}^{\prime }\right) ,
\end{equation}%
we apply the functional differentiation to obtain:
\begin{eqnarray}
0 &=&\frac{\delta }{\delta H\left( \mathbf{r}^{\prime }\right) }\left[
-\gamma \nabla ^{2}L\left( \mathbf{r}\right) -\rho \left( L\left( \mathbf{r}%
\right) \right) -H\left( \mathbf{r}\right) \right]  \notag \\
&=&\frac{1}{k_{B}T}\left[ -\gamma \nabla ^{2}-\frac{\partial \rho \left(
L\left( \mathbf{r}\right) \right) }{\partial L\left( \mathbf{r}\right) }%
\right] G\left( \mathbf{r,r}^{\prime }\right) -\delta \left( \mathbf{r}-%
\mathbf{r}^{\prime }\right) .  \notag \\
&&
\end{eqnarray}%
which is equivalent to Eqn.~(\ref{Green}). It is clear that $G\left( \mathbf{%
r,r}^{\prime }\right) $ is the propagator of Eqn.\ (\ref{eqmotion}).
\begin{figure}[tb]
\center{\includegraphics[width=8cm]{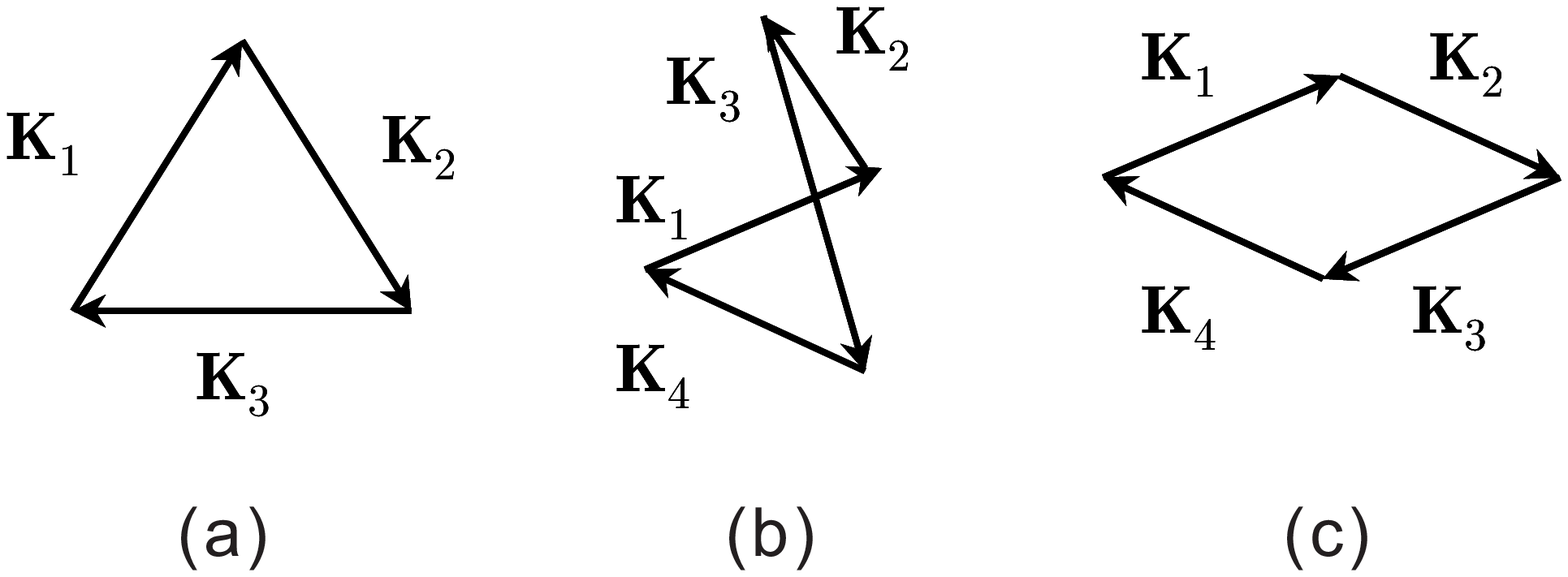}}
\caption{(a) $\mathbf{K}_{1}+\mathbf{K}_{2}+\mathbf{K}_{3}=0$ where $%
\left\vert \mathbf{K}_{1}\right\vert =\left\vert \mathbf{K}_{2}\right\vert
=\left\vert \mathbf{K}_{3}\right\vert $. (b) $\mathbf{K}_{1}+\mathbf{K}_{2}+%
\mathbf{K}_{3}+\mathbf{K}_{4}=0$ where $\left\vert
\mathbf{K}_{1}\right\vert =\left\vert \mathbf{K}_{2}\right\vert
=\left\vert \mathbf{K}_{3}\right\vert =\left\vert
\mathbf{K}_{4}\right\vert $. In general, such \textquotedblleft
off-plane\textquotedblright\ term $\mu \left(
\mathbf{K}_{1}\right)
\protect\mu \left( \mathbf{K}_{2}\right) \protect\mu \left( \mathbf{K}%
_{3}\right) \protect\mu \left( \mathbf{K}_{4}\right) $ does not contain $%
\left\vert \protect\mu \left( \mathbf{K}_{1}\right) \right\vert ^{2}$
explicitly. (c) \textquotedblleft In-plane\textquotedblright\ four wave
vectors. In such case, $\protect\mu \left( \mathbf{K}_{1}\right) \protect\mu %
\left( \mathbf{K}_{2}\right) \protect\mu \left( \mathbf{K}_{3}\right)
\protect\mu \left( \mathbf{K}_{4}\right) =\left\vert \protect\mu \left(
\mathbf{K}_{1}\right) \right\vert ^{2}\left\vert \protect\mu \left( \mathbf{K%
}_{2}\right) \right\vert ^{2}$.}
\label{FIG5:e}
\end{figure}
\begin{figure}[tb]
\center{\includegraphics[width=8cm]{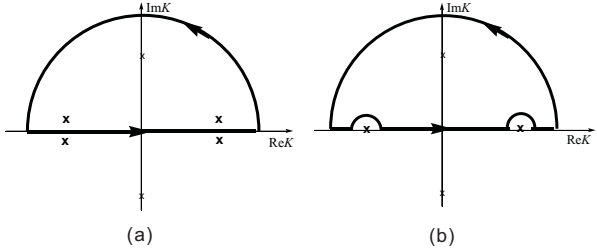}} \caption{(a) This
plots the poles (x) and the integration contour (thick lines) for
$M_{\protect\varepsilon }(K)$ in the complex $K$ plane. According
to Jordan's lemma, only the residues at the poles in the upper
plane are taken into account. (b) This plots the poles (x) and the
integration contour
(thick lines) for $M(K)$. Note that the 4 poles of $M_{\protect%
\varepsilon }(K)$ near the real axis are merged as 4/2=2 poles in $M\left( K\right) $%
, so every pole on the real axis is wound around only half a
circle in the contour designed for $M\left( K\right) $.}
\label{FIG6:f}
\end{figure}

For $T\ll T_{m}$, where $T_{m}$ is defined by Eqn.~(\ref{Tm}), $\rho \left(
L\left( \mathbf{r}\right) \right) $ is almost linear in the vicinity of $%
\left\langle L\right\rangle _{T}$, so $-\partial \rho \left( L\left( \mathbf{%
r}\right) \right) /\partial L\left( \mathbf{r}\right) $ is almost a constant
$\rho _{1}$ even when fluctuation is considered. The solution of the
propagator is then
\begin{eqnarray}
G\left( \mathbf{r,r}^{\prime }\right) &=&\frac{k_{B}T}{4\pi \gamma r}e^{-r/%
{\xi^*_{T}}},r=\left\vert \mathbf{r-r}^{\prime }\right\vert ,
\notag \\
{\xi^*_{T} } &=&\sqrt{\frac{\gamma }{\rho _{1}}}.
\end{eqnarray}%
However, for $T\rightarrow T_{m}-0$, the linearity of $\rho \left( L\left(
\mathbf{r}\right) \right) $ in the vicinity of $\left\langle L\right\rangle
_{T}$ is lost and the corresponding non-Gaussian fluctuations of $L\left(
\mathbf{r}\right) $ in different magnitudes have different propagation modes
and correlation length. A fluctuation characterized by $\left\vert L\left(
\mathbf{r}\right) \right\vert >\left\vert \left\langle L\right\rangle
_{T_{m}}\right\vert $ decays exponentially (because $\partial \rho \left(
L\left( \mathbf{r}\right) \right) /\partial L\left( \mathbf{r}\right) <0$ in
this case), but fluctuations with $\left\vert L\left( \mathbf{r}\right)
\right\vert <\left\vert \left\langle L\right\rangle \right\vert _{T_{m}}$%
could be propagated in the form of a sinusoidal wave (because $\partial \rho
\left( L\left( \mathbf{r}\right) \right) /\partial L\left( \mathbf{r}\right)
>0$ in that case). The mean wavelength will be estimated in the next
subsection.

Such fluctuation propagation is a consequence of plasma instability at $%
T_{m} $, and gives rise to collaborative atom clusters, the shape of which
is nearly spherical, for the reason outlined in the next subsection.

\subsection{The Correlation Functions When $T\rightarrow T_{m}-0$}

The mean wavelength of $L\left( \mathbf{r}\right) $ fluctuation
correlation could be estimated with the following steps:

\textit{Step 1:} Define the Fourier expansion of $\mu \left(
\mathbf{r}\right) $ as
\begin{equation}
\mu \left( \mathbf{r}\right) =\sum_{\mathbf{K}}\mu \left( \mathbf{K}\right)
e^{i\mathbf{K}\cdot \mathbf{r}}
\end{equation}%
so that the complex conjugate of $\mu \left( \mathbf{K}\right) $ is $\mu
\left( -\mathbf{K}\right) $ and that
\begin{equation}
\nabla \mu \left( \mathbf{r}\right) =\sum_{\mathbf{K}}i\mathbf{K}\mu \left(
\mathbf{K}\right) e^{i\mathbf{K}\cdot \mathbf{r}}.
\end{equation}%
Expand the free energy functional in the power series of $\mu \left( \mathbf{%
K}\right) $ in a volume $V=2N$ which contains $2N$ sites:
\begin{widetext}
\begin{eqnarray}
&&F-F\left[ \left\langle L\right\rangle _{T_{m}}\right] \notag \\&=& \int \mathrm{d}^{3}%
\mathbf{r\ }\left\{ \frac{1}{3}\mu ^{3}\left( \mathbf{r}\right) \rho _{2}+%
\frac{1}{4}\mu ^{4}\left( \mathbf{r}\right) \rho _{3}+\frac{\gamma }{2}\left[
\nabla \mu \left( \mathbf{r}\right) \right] ^{2}\right\}   \notag \\
&=&\int \mathrm{d}^{3}%
\mathbf{r\ \ }\left\{ \frac{\gamma}{2}\sum_{\mathbf{K,K}^{\prime }}(i%
\mathbf{K})(i\mathbf{K}^{\prime })\mu \left( \mathbf{K}\right) \mu \left(
\mathbf{K}^{\prime }\right) e^{i\left( \mathbf{K+K}^{\prime }\right) \cdot
\mathbf{r}}+\frac{\rho _{2}}{3}\sum_{\mathbf{K}_{1}\mathbf{,K}_{2},\mathbf{K}%
_{3}}\mu \left( \mathbf{K}_{1}\right) \mu \left( \mathbf{K}_{2}\right) \mu
\left( \mathbf{K}_{3}\right) e^{i\left( \mathbf{K}_{1}\mathbf{+K}_{2}+%
\mathbf{K}_{3}\right) \cdot \mathbf{r}}\right.   \notag \\
&&\left. +\frac{\rho _{3}}{4}\sum_{\mathbf{K}_{1}\mathbf{,K}_{2},\mathbf{K}%
_{3},\mathbf{K}_{4}}\mu \left( \mathbf{K}_{1}\right) \mu \left( \mathbf{K}%
_{2}\right) \mu \left( \mathbf{K}_{3}\right) \mu \left( \mathbf{K}%
_{4}\right) e^{i\left( \mathbf{K}_{1}\mathbf{+K}_{2}+\mathbf{K}_{3}+\mathbf{K%
}_{4}\right) \cdot \mathbf{r}}\right\}   \notag \\
&=&V\sum_{\mathbf{K}}\frac{\gamma}{2}K^{2}\left\vert \mu \left(
\mathbf{K}\right) \right\vert
^{2}+\frac{V\rho _{2}}{3}\sum_{\mathbf{K}_{1}+\mathbf{K}_{2}+\mathbf{K}%
_{3}=0}\mu \left( \mathbf{K}_{1}\right) \mu \left( \mathbf{K}_{2}\right) \mu
\left( \mathbf{K}_{3}\right)+\frac{V\rho _{3}}{4}\sum_{\mathbf{K}_{1}+\mathbf{K}_{2}+\mathbf{K}_{3}+%
\mathbf{K}_{4}=0}\mu \left( \mathbf{K}_{1}\right) \mu \left( \mathbf{K}%
_{2}\right) \mu \left( \mathbf{K}_{3}\right) \mu \left( \mathbf{K}%
_{4}\right) \notag \\
&& \label{Ffluc}
\end{eqnarray}%
where $K=\left\vert \mathbf{K}\right\vert $.

\textit{Step 2:} Evaluate $\langle \left\vert \mu \left(
\mathbf{K}\right) \right\vert ^{2}\rangle $ by
\begin{equation}
\left\langle \left\vert \mu \left( \mathbf{K}\right) \right\vert
^{2}\right\rangle =\left. \int \left\vert \mu \left( \mathbf{K}\right)
\right\vert ^{2}e^{-\frac{F}{k_{B}T_{m}}}\prod_{\mathbf{K}}\mathrm{d}\mu
\left( \mathbf{K}\right) \right/ \int e^{-\frac{F}{k_{B}T_{m}}}\prod_{%
\mathbf{K}}\mathrm{d}\mu \left( \mathbf{K}\right)
\end{equation}

Based on the assumption that the fluctuation wave has a definite
wave length $2\pi \xi $, we can assert that only $\left\vert
\mathbf{K}\right\vert
\approx 1/\xi $ terms dominate the sum in Eqn.~(\ref{Ffluc}) (FIG.~\ref%
{FIG5:e} and Ref.~\cite{Bak}) and carry out the summation only on
equilateral polygons of $\mathbf{K}$. This results in
\begin{eqnarray}
&&F-F\left[ \left\langle L\right\rangle _{T_{m}}\right]   =V\sum_{\mathbf{K}}\frac{\gamma }{2}K^{2}\left\vert \mu \left( \mathbf{K}%
\right) \right\vert ^{2}+O\left( \mu ^{3}\right)   +\frac{V\rho
_{3}}{4}\sum_{\left\vert \mathbf{K}\right\vert =\left\vert
\mathbf{K}^{\prime }\right\vert \approx \xi ^{-1}}\left\vert \mu
\left( \mathbf{K}\right) \right\vert ^{2}\left\vert \mu \left(
\mathbf{K}^{\prime }\right) \right\vert ^{2}
\end{eqnarray}%
where \textquotedblleft off-plane\textquotedblright\ contributions of $\mu
\left( \mathbf{K}_{1}\right) $ $\mu \left( \mathbf{K}_{2}\right) $ $\mu
\left( \mathbf{K}_{3}\right) $ $\mu \left( \mathbf{K}_{4}\right) $ (FIG.~%
\ref{FIG5:e} (b)) are assumed to be cancelled by phase randomness. By using
the density of states in $\mathbf{K}$-space, one may find %
\begin{eqnarray}
(i) &&\sum_{\left\vert \mathbf{K}^{\prime }\right\vert \approx \xi
^{-1}}\left\vert \mu \left( \mathbf{K}^{\prime }\right)
\right\vert ^{2} =\frac{V}{\left( 2\pi \right) ^{3}}\int_{\xi
^{-1}-\Delta K}^{\xi ^{-1}+\Delta K}\left\vert \mu \left(
\mathbf{K}^{\prime }\right) \right\vert
^{2}4\pi K^{2}dK  \notag \\
&\approx &\frac{V}{6\pi ^{2}}\left\vert \mu \left(
\mathbf{K}^{\prime }\right) \right\vert ^{2}\left[ \left( \xi
^{-1}+\Delta K\right) ^{3}-\left( \xi ^{-1}-\Delta K\right)
^{3}\right] =\frac{Vc\left( c^{2}+3\right) }{3\pi
^{2}}\frac{\left\vert \mu \left( \mathbf{K}^{\prime }\right)
\right\vert ^{2}}{\xi ^{3}}
\end{eqnarray}
when the \textquotedblleft spectral width\textquotedblright\ reads
$\Delta K=c\xi ^{-1}$;
\begin{eqnarray}
(ii) &&\sum_{\left\vert \mathbf{K}\right\vert =\left\vert
\mathbf{K}^{\prime }\right\vert \approx \xi ^{-1}}\left\vert \mu
\left( \mathbf{K}\right) \right\vert ^{2}\left\vert \mu \left(
\mathbf{K}^{\prime }\right) \right\vert ^{2} \approx
\frac{Vc\left( c^{2}+3\right) }{3\pi ^{2}\xi ^{3}}\sum_{\left\vert
\mathbf{K}\right\vert \approx \xi ^{-1}}\left\vert \mu \left( \mathbf{K}%
\right) \right\vert ^{4},
\end{eqnarray}%
which leads to the approximation:
\begin{eqnarray}
&&\int \mathrm{d}\mu \left( \mathbf{K}\right) \ \exp \left\{ -\frac{V}{%
k_{B}T_{m}}\left[ \frac {\gamma}{2} K^{2}\left| \mu \left( \mathbf{K}%
\right) \right| ^{2}+O\left( \mu ^{3}\right) +\frac{\rho _{3}}{4}\frac{%
Vc\left( c^{2}+3\right) }{3\pi ^{2}\xi ^{3}}\left| \mu \left( \mathbf{K}%
\right) \right| ^{4}\right] \right\}   \notag \\
&=&\int \mathrm{d}\mu \left( \mathbf{K}\right) \exp \left\{
-\frac{V}{ k_{B}T_{m}}\left[\frac {\gamma}{2} K^{2}\left| \mu
\left( \mathbf{K}\right)
\right| ^{2}\right] \right\} \left[1-\frac{V\rho _{3}}{4k_{B}T_{m}}\frac{%
Vc\left( c^{2}+3\right) }{3\pi ^{2}\xi ^{3}}\left| \mu \left( \mathbf{K}%
\right) \right| ^{4}+O\left( \mu ^{6}\right) \right]  \notag \\
&=&\left( \frac{2k_{B}T_{m}}{V\gamma K^{2}}\right) ^{1/2}-\frac{V\rho _{3}}{%
4k_{B}T_{m}}\frac{Vc\left( c^{2}+3\right) }{3\pi ^{2}\xi ^{3}}\frac{3}{4}%
\left( \frac{2k_{B}T_{m}}{V\gamma K^{2}}\right) ^{5/2}+O\left( \frac{1}{%
K^{7}}\right)
\end{eqnarray}%
\begin{eqnarray}
& &\int \mathrm{d}\mu \left( \mathbf{K}\right) \ \left| \mu \left( \mathbf{K}%
\right) \right| ^{2}\exp \left\{ -\frac{V}{k_{B}T_{m}}\left[ \frac{\gamma }{%
2}K^{2}\left| \mu \left( \mathbf{K}\right) \right| ^{2}+O\left( \mu
^{3}\right) +\frac{\rho _{3}}{4}\frac{Vc\left( c^{2}+3\right) }{3\pi ^{2}\xi
^{3}}\left| \mu \left( \mathbf{K}\right) \right| ^{4}\right] \right\}
\notag \\
&=&\int \mathrm{d}\mu \left( \mathbf{K}\right) \left[ \left| \mu
\left( \mathbf{K}\right) \right| ^{2}+O\left( \mu ^{6}\right)
\right] \exp \left\{
-\frac{V}{k_{B}T_{m}}\left[ \frac{\gamma }{2}K^{2}\left| \mu \left( \mathbf{K%
}\right) \right| ^{2}\right] \right\}   \notag \\
&=&\frac{1}{2}\left( \frac{2k_{B}T_{m}}{V\gamma K^{2}}\right)
^{3/2}\left[ 1+O\left( \frac{1}{K^{4}}\right) \right]
\end{eqnarray}%
for short range correlations where the $K^{2}\mu ^{2}$ term overwhelms the $%
\mu ^{4}$ term, and this finally leads to
\begin{eqnarray}
\langle \left\vert \mu \left( \mathbf{K}\right) \right\vert ^{2}\rangle  &=&%
\frac{1}{2}\left( \frac{2k_{B}T_{m}}{V\gamma K^{2}}\right) ^{3/2}{\left[
\left( \frac{2k_{B}T_{m}}{V\gamma K^{2}}\right) ^{1/2}-\frac{V\rho _{3}}{%
4k_{B}T_{m}}\frac{Vc\left( c^{2}+3\right) }{3\pi ^{2}\xi ^{3}}\frac{3}{4}%
\left( \frac{2k_{B}T_{m}}{V\gamma K^{2}}\right) ^{5/2}+O\left( \frac{1}{%
K^{7}}\right) \right] }^{-1}  \notag \\
&=&\frac{1}{2}\frac{V\gamma K^{2}}{2k_{B}T_{m}}\left[ \left( \frac{%
V\gamma K^{2}}{2k_{B}T_{m}}\right) ^{2}-\frac{V^{2}\rho _{3}}{16k_{B}T_{m}}%
\frac{c\left( c^{2}+3\right) }{\pi ^{2}\xi ^{3}}+O\left( \frac{1}{K^{2}}%
\right) \right] ^{-1}  \notag \\
&=&\frac{k_{B}T_{m}\gamma K^{2}}{V}\left[ \gamma ^{2}K^{4}-\frac{1}{4}\frac{%
\rho _{3}k_{B}T_{m}}{\pi ^{2}\xi ^{3}}c\left( c^{2}+3\right) +O\left( \frac{1%
}{K^{2}}\right) \right] ^{-1}.
\end{eqnarray}%
Let
\begin{eqnarray}
M_{\varepsilon }\left( K\right) &=&{\gamma K^{2}}\left[\gamma ^{2}K^{4}-%
\frac{1}{4}\frac{\rho _{3}k_{B}T_{m}}{\pi ^{2}\xi ^{3}}c\left(
c^{2}+3\right) +O\left( \frac{1}{K^{2}}\right) \right]^{-1}, \\
M\left( K\right) &=&{\gamma K^{2}}\left[\gamma ^{2}K^{4}-\frac{1}{4}\frac{%
\rho _{3}k_{B}T_{m}}{\pi ^{2}\xi ^{3}}c\left( c^{2}+3\right)
\right]^{-1}.
\end{eqnarray}%
where $K$ is a complex variable in general. $M_{\varepsilon
}\left( K\right) $ has six poles \cite{posdef}. Three of these
poles lie above the real axis, and the other three conjugate poles
lie below the real axis. $M\left( K\right) $, which is an
approximation of $M_{\varepsilon }\left( K\right) $, has four
poles (FIG.~(\ref{FIG6:f})). Among the four, there are two poles
on the real $K$ axis corresponding to the four poles of
$M_{\varepsilon }\left( K\right) $ that lie near the real axis.
Each of the two poles should be regarded as half a pole that
corresponds to the contribution from one upper-plane pole in
$M_{\varepsilon }\left( K\right) $ when applying the residue
theorem to evaluate the contour integral. Therefore, the
coefficient before the summation of the residues at these two
poles is $\pi i$ instead of $2\pi i$ in the equation below. The
two real-valued poles of $M\left( K\right) $ contribute to a
sinusoidal wave. The Green function corresponding to this
non-Gaussian fluctuation is then
\begin{eqnarray}
G\left( r\right) &=&\frac{V}{\left( 2\pi \right) ^{3}}\int \mathrm{d}^{3}%
\mathbf{K\ }\langle \left| \mu \left( \mathbf{K}\right) \right|
^{2}\rangle e^{i\mathbf{K}\cdot \mathbf{r}}  \notag \\
&=&\frac{V}{\left( 2\pi \right) ^{3}}\int_{0}^{+\infty }K^{2}\mathrm{d}%
K\int_{0}^{\pi }e^{iKr\cos \theta }\sin \theta \mathrm{d}\theta
\int_{0}^{2\pi }\mathrm{d}\phi \ \langle \left| \mu \left( \mathbf{K}%
\right) \right| ^{2}\rangle  \notag \\
&=&\frac{k_{B}T_{m}}{2\pi ^{2}r}\int_{0}^{+\infty }M_{\varepsilon
}\left(
K\right) K\sin Kr\mathrm{d}K  \notag \\
&=&-\frac{k_{B}T_{m}}{4\pi ^{2}r}\frac{\partial }{\partial
r}\int_{-\infty
}^{+\infty }M_{\varepsilon }\left( K\right) \cos Kr\mathrm{d}K  \notag \\
&=&-\frac{k_{B}T_{m}}{4\pi ^{2}r}\frac{\partial }{\partial
r}\left[ 2\pi
i\sum_{\mathrm{Im}K>0}\text{res}(M_{\varepsilon }\left( K\right) e^{iKr},K)%
\right]  \notag \\
&\approx &-\frac{k_{B}T_{m}}{4\pi ^{2}r}\frac{\partial }{\partial
r}\left[
2\pi i\sum_{\mathrm{Im}K>0}\text{res}(M\left( K\right) e^{iKr},K)+\pi i\sum_{%
\mathrm{Im}K=0}\text{res}(M\left( K\right) e^{iKr},K)\right]  \notag \\
&=&\frac{k_{B}T_{m}}{8\pi \gamma r}\left( e^{-K_{0}r}+\cos
K_{0}r\right) .
\end{eqnarray}
\end{widetext}Here,
\begin{equation}
K_{0}^{4}=\frac{1}{\xi ^{4}}=\frac{1}{4}\frac{\rho _{3}k_{B}T_{m}}{\pi
^{2}\gamma ^{2}\xi ^{3}}c\left( c^{2}+3\right)
\end{equation}%
which means
\begin{equation*}
\xi =\frac{4\pi ^{2}\gamma ^{2}2^{3/2}}{\rho _{3}k_{B}T_{m}c\left(
c^{2}+3\right) a^{3}}\geq \frac{2\sqrt{2}\pi ^{2}\gamma ^{2}}{\rho
_{3}k_{B}T_{m}a^{3}}
\end{equation*}%
where the NNN atom distance $a$ is restored. (The inequality
results from the fact that $c\leq1$ -- that is, the maximum
spectral width cannot exceed $K_0$.)

Therefore,
\begin{equation}
\pi \xi _{\min }=\frac{2\sqrt{2}\pi ^{3}\gamma ^{2}}{\rho _{3}k_{B}T_{m}a^{3}%
}  \label{ksimin}
\end{equation}%
gives an estimate of the minimal diameter of unstable clusters in
the solid when $T=T_{m}$.

\acknowledgements This work is supported by National Natural Science
Foundation of China and 973 project.


\begin{thebibliography}{99}
\bibitem{Cahn} R. W. Cahn, Nature (London) \textbf{273}, 491 (1978);\textbf{%
323}, 668 (1986)

\bibitem{rev} J. G. Dash, Rev. Mod. Phys. \textbf{71}, 1737 (1999)

\bibitem{Lindemann} F. A. Lindemann, Phys. Zeit. \textbf{11}, 609 (1910)%
\newline
J. J. Gilvarry, Phys. Rev. \textbf{102}, 308 (1956)

\bibitem{Born} M. Born, J. Chem. Phys. \textbf{7}, 591 (1939)

\bibitem{Jin} Z. H. Jin, P. Gumbsch, K. Lu and E. Ma, Phys. Rev. Lett.
\textbf{87}, 055703 (2001)

\bibitem{Brule} J. L. Tallon, Phil. Mag. A \textbf{39}, 151 (1978)

\bibitem{Lrule} G. Grimvall and S. Sjodin, Physica Scripta, \textbf{10}, 340
(1974)

\bibitem{surface melting} A. Trayanov and E. Tosatti, Phys. Rev. B, \textbf{%
38}, 6961 (1988); R. Ohnesorge, H. L\"{o}wen and H. Wagner, Phys. Rev. E,
\textbf{50}, 4801 (1994); J. Q. Broughton and G. H. Gilmer, J. Chem. Phys.
\textbf{79}, 5095 (1983); \textbf{79}, 5105 (1983)

\bibitem{Stilinger and Weber} F. H. Stillinger and T. A. Weber, J. Chem.
Phys. \textbf{81}, 5095 (1984)

\bibitem{first-order transition} J. Cardy, J. Phys.\ A: Math. Gen. \textbf{29%
}, 1897 (1996)

\bibitem{Granato} A. V. Granato, Phys. Rev. Lett. \textbf{68}, 974 (1992)

\bibitem{Baskes} M. I. Baskes, Phys. Rev. Lett. \textbf{83}, 2592 (1999)

\bibitem{Halpern} V. Halpern, J. Phys.:Condens. Matter \textbf{12}, 4303
(2000)

\bibitem{Sandberg} N. Sandberg, B. Magyari-K\"{o}pe, and T. R. Mattsson,
Phys. Rev. Lett. \textbf{89}, 065901 (2002)

\bibitem{Self-interstitial} K. Nordlund and R. S. Averback, Phys. Rev. Lett.
\textbf{80}, 4201 (1998)

\bibitem{Si-interstitial} A. Bongiorno, L. Colombo, F. Cargnoni, C. Gatti
and M. Rosati, Europhys. Lett., \textbf{50}, 608 (2000)

\bibitem{Ge-interstitial} S. Birner, J. P. Goss and R. Jones, P. R. Briddon,
S. \"{O}berg, \textit{Proceedings of ENDEASD }(Stockholm 2000)

\bibitem{As-interstitial} J. I. Landman and C. G. Morgan, J. T. Schick, P.
Papoulias and A. Kumar, Phys. Rev. B \textbf{55}, 15\thinspace 581 (1997)

\bibitem{Juelich} K. W. Ingle, R. C. Perrin and H. R. Schober, J. Phys. F:
Metal Phys, \textbf{11}, 1161 (1981)

\bibitem{deffreq} L. Salter, Proc. Roy. Soc. (London) \textbf{A233}, 418
(1956); L. Dobrzynski and J. Friedel, Surf. Science \textbf{12}, 649 (1968);
G. Allan and M. Lannoo, in \textit{Defects and Radiation Effects in
Semiconductors} (ed. R. R. Hasiyuti), The Institute of Physics (Bristol and
London), 1980

\bibitem{Landau} L. D. Landau and E. M. Lifshitz, \textit{Statistical
Mechanics, Part I} (Butterworth-Heinemann, 1999)

\bibitem{vol} For the moment, we also neglect the thermal expansion of the
solid, which we do not regard as the decisive driving force of melting. This
is based on the commonsense that solid does not necessarily
\textquotedblleft expand to melt\textquotedblright . H$_{2}$O and Bi are two
telling examples in which the liquid phase is even more dense than the solid
phase. The densities of ice and water are $0.9\times 10^{3}\mathrm{kg\,m}%
^{-3}$ and $1\times 10^{3}\mathrm{kg\,m}^{-3}$ respectively. The density of
solid bismuth is $9747\mathrm{kg\,m}^{-3}$, less than that of the liquid
counterpart: $10050\mathrm{kg\,m}^{-3}$.

\bibitem{Pathria} R. K. Pathria, \textit{Statistical Mechanics}, 2nd ed.,
Elsevier (Singapore) Pte Ltd, 2001

\bibitem{Huang} Kerson Huang, \textit{Introduction to Statistical Physics},
Taylor \& Francis, 2001

\bibitem{Yang and Lee} C. N. Yang and T. D. Lee, Phys. Rev. \textbf{87}, 404
(1957)\newline
T. D. Lee and C. N. Yang, Phys. Rev. \textbf{87}, 410 (1957).

\bibitem{Frustrated magnets} A. F. Barabanov and V. M. Berezovski\u{\i}, Zh.
Eksp. Teor. Fiz. \textbf{106}, 1156 (1994); H. Rosner, R. Hayn and J.
Schulenburg, Phys. Rev. B \textbf{57}, 13660 (1998); J. Richter, A. Voigt,
J. Schulenburg, N. B. Ivanoc and R. Hayn, J. Magn. Magn. Mat. \textbf{177-181%
}, 737 (1998)

\bibitem{Nigel Goldenfeld} N. Goldenfeld, \textit{Lectures on Phase
Transitions and the Renormalization Group} (Addison Wesley, 1992)

\bibitem{Ginzburg} V. L. Ginzburg, Sov. Phys. Solid State \textbf{2}, 1824
(1960)

\bibitem{Jackson} J. D. Jackson, \textit{Classical Electrodynamics}, 3rd
ed., Wiley Text Books, 1998

\bibitem{dumbbell} Previous studies in interstitial clusters (\textit{e.g.}~Refs. \cite{Juelich, Si-interstitial}) also
showed that the excitation energy for a di-interstitial with the
``dumbbell'' configuration is less than that of two split
interstitials. In other words, previous studies have shown that it
costs less than twice the excitation energy of a single isolated
interstitial to excite two interstitial defects that are in close
contact.

\bibitem{hconc} When defect concentration is $1/24\approx 0.042$, roughly
speaking, the $12$ NNN sites adjacent to one atom report half an
atom missing. The absence of half an atom in the next nearest
neighbor site could make enough room for the transference of the
energy and information between atoms.
Such transference is responsible for the propagation of the fluctuation wave. Therefore, $%
3\%\sim 5\%$ concentration is also \textquotedblleft high
enough\textquotedblright .

\bibitem{Bak} P. Bak, Phys. Rev. Lett. \textbf{54}, 1517 (1985)

\bibitem{posdef} By the positive-definiteness of
$\langle \left\vert \mu \left( \mathbf{K}\right) \right\vert
^{2}\rangle$, it is easy to see that the function $M_{\varepsilon
}\left( K\right) $ is greater than zero for all real $K$ values.
However, the function $M(K)$ is not positive-definite for all real
$K$ values. Nevertheless, this disadvantageous property of $M(K)$
does not affect the evaluation of the integral.
\end{thebibliography}
\end{document}